\documentclass[journal]{IEEEtran}
\usepackage{graphicx}
\usepackage{textcomp}
\usepackage{xcolor}
\usepackage{colortbl}
\definecolor{mygray}{gray}{0.98}
\definecolor{mygray1}{gray}{0.93}
\usepackage{algorithm}  
\usepackage{algpseudocode}  
\usepackage{amsmath}  
\usepackage{amsthm}
\usepackage{amsfonts}
\usepackage{amssymb}
\usepackage{latexsym}
\usepackage{caption, subcaption}
\usepackage{enumitem}
\usepackage{booktabs}
\usepackage{listings}
\usepackage{chngpage}
\usepackage{flushend}
\usepackage{makecell}
\usepackage{adjustbox}
\usepackage{bbding}
\usepackage{utfsym}
\usepackage{autobreak}
\usepackage{multirow}
\usepackage{siunitx}

\usepackage{textcomp,booktabs}
\usepackage{wasysym}
\newcommand{\circled}[1]{\normalsize{\textcircled{\scriptsize{#1}}}\normalsize\;}
\usepackage{ragged2e}
\definecolor{seagreen}{rgb}{0.18, 0.55, 0.34}
\definecolor{royalpurple}{rgb}{0.47,0.32,0.66}
\definecolor{brown(traditional)}{rgb}{0.59, 0.29, 0.0}
\definecolor{blue}{rgb}{0.3, 0.2, 0.9}
\usepackage[colorlinks,
            linkcolor=blue,
            anchorcolor=blue,
            citecolor=blue]{hyperref}
\ifCLASSINFOpdf
\else
\fi
\begin{document}
%
\title{Intelligent Mobile AI-Generated Content Services via Interactive Prompt Engineering and Dynamic Service Provisioning}
%
%
%

\author{Yinqiu~Liu,
        Ruichen~Zhang,
        Jiacheng~Wang,
        Dusit~Niyato,~\IEEEmembership{Fellow,~IEEE},\\
        Xianbin~Wang,~\IEEEmembership{Fellow,~IEEE},
        Dong~In~Kim,~\IEEEmembership{Life Fellow,~IEEE},
        and Hongyang~Du
\thanks{Y.~Liu, R.~Zhang, J.~Wang, and D.~Niyato are with the College of Computing and Data Science, Nanyang Technological University, Singapore (e-mails: yinqiu001@e.ntu.edu.sg, ruichen.zhang@ntu.edu.sg, jiacheng.wang@ntu.edu.sg, and dniyato@ntu.edu.sg).}
\thanks{X.~Wang is with the Department of Electrical and Computer Engineering, Western University, Canada (e-mail: xianbin.wang@uwo.ca).}
\thanks{D.~Kim is with the College of Information and Communication Engineering, Sungkyunkwan University, South Korea (e-mail: dongin@skku.edu).}
\thanks{H.~Du is with the Department of Electrical and Electronic Engineering, University of Hong Kong, Hong Kong SAR, China (e-mail: duhy@eee.hku.hk).}
\vspace{-0.15cm}
}

\maketitle

\begin{abstract}
Due to massive computational demands of large generative models, AI-Generated Content (AIGC) can organize collaborative Mobile AIGC Service Providers (MASPs) at network edges to provide ubiquitous and customized content generation for resource-constrained users. 
However, such a paradigm faces two significant challenges: i) raw prompts (i.e., the task description from users) often lead to poor generation quality due to users' lack of experience with specific AIGC models, and ii) static service provisioning fails to efficiently utilize computational and communication resources given the heterogeneity of AIGC tasks. To address these challenges, we propose an intelligent mobile AIGC service scheme. Firstly, we develop an interactive prompt engineering mechanism that leverages a Large Language Model (LLM) to generate customized prompt corpora and employs Inverse Reinforcement Learning (IRL) for policy imitation through small-scale expert demonstrations. Secondly, we formulate a dynamic mobile AIGC service provisioning problem that jointly optimizes the number of inference trials and transmission power allocation. Then, we propose the Diffusion-Enhanced Deep Deterministic Policy Gradient (D$^3$PG) algorithm to solve the problem. By incorporating the diffusion process into Deep Reinforcement Learning (DRL) architecture, the environment exploration capability can be improved, thus adapting to varying mobile AIGC scenarios. Extensive experimental results demonstrate that our prompt engineering approach improves single-round generation success probability by 6.3$\times$, while D$^3$PG increases the user service experience by 67.8\% compared to baseline DRL approaches.
\end{abstract}

\begin{IEEEkeywords}
Mobile AI-generated content, prompt engineering, large language model, inverse reinforcement learning
\end{IEEEkeywords}

%
\IEEEpeerreviewmaketitle

\section{Introduction}
%
%
%
%
\IEEEPARstart{R}{ecently}, AI-Generated Content (AIGC) \cite{10398474, DuAIGC} has sparked significant interest across both academic and industrial sectors. 
Notable AIGC tools along this trend include DALL$\cdot$E 3, MusicLM, and ChatGPT for image generation, music composition, and multimodal conversation, respectively \cite{YQNetwork}. 
However, such achievements are built on large foundation models comprising massive parameters.
For example, GPT-3, released in 2020, already contains 175 billion parameters.
Accordingly, training such a model on a single GPU takes 355 years and consumes \$4.6 million \cite{GPT-3Cost}.
However, hardware scaling has not kept pace with the explosion in model parameter volume and resource requirements.
As the latest mobile AI chip, \textit{Qualcomm Snapdragon 8 Gen 3} can only afford lightweight AIGC models with roughly ten billion parameters \cite{Snapdragon}.
Constrained by Moore's law, it is foreseeable that such lightweight AIGC models will still be the mainstream for mobile deployment over a long period.
The conflict between model overhead and hardware capabilities prevents users from using ubiquitous high-quality AIGC services.

To address this challenge, the concept of \textit{Mobile AIGC} has been proposed, utilizing mobile-edge computing to democratize high-quality AIGC services \cite{10398474, 10628024}. 
Specifically, resource-constrained mobile users delegate their AIGC tasks to Mobile AIGC Service Providers (MASPs) served by edge servers, base stations, etc. \cite{10398474}. 
These MASPs, equipped with sufficient computational power, perform generative inferences, offering on-demand and paid AIGC services based on users' requirements (so-called prompts). 
This approach can not only alleviate the computational burden on individual users but also enhance privacy by reducing the need to send sensitive information to distant cloud servers \cite{10398474}. 
Great efforts in terms of model and networking have been made to promote the development of mobile AIGC.
For instance, Qualcomm \cite{qualcomm}, Salimans \textit{et al.} \cite{knowledge}, and Chen \textit{et al.} \cite{chen2023speed} adopted quantization, knowledge distillation, and GPU-aware optimization to compress AIGC models, respectively.
From the network perspective, Xu \textit{et al.} \cite{xu2023sparks} optimized the caching strategy in mobile AIGC, facilitating MASPs to manage their local AIGC models efficiently.
Additionally, Du \textit{et al.} \cite{10172151} presented a distributed manner of mobile AIGC inference, realizing the customized and collaborative AIGC generations.
Wen \textit{et al.} \cite{10233667} scheduled the task allocation among multiple MASPs and optimized the incentive mechanism to encourage them to invest computation resources.
\begin{figure}[tbp]
\centerline{\includegraphics[width=0.99\columnwidth]{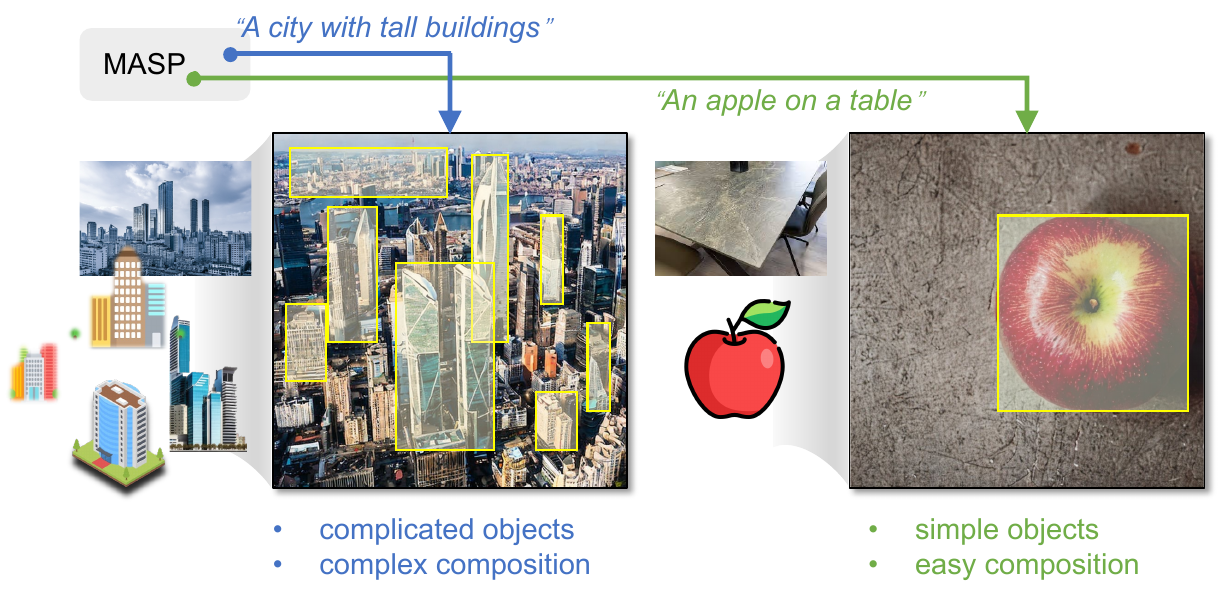}}
\caption{The heterogeneity of AIGC tasks. We can obverse that generating an image of a city is much more difficult than that of an apple since more complicated objects and compositions are required. Therefore, more inference trials should be allocated. Moreover, complex images accommodate more information (e.g., edges and visual signals) \cite{Complexity}. Hence, they are more sensitive to transmission loss and require more transmission power.}
\vspace{-0.12cm}
\label{example}
\end{figure}

Despite such progress, existing mobile AIGC schemes all follow a basic service paradigm, i.e., mobile users upload their prompts, and the MASPs perform AIGC inferences accordingly \cite{10398474, DuAIGC, 10628024, 10233667}.
We can observe that two challenges exist in this process.
\begin{itemize}
    \item \textbf{Low-Quality Raw Prompts:} As the description of user requirements and the instruction for AIGC inferences, prompts directly determine generation quality. Unfortunately, existing proposals \cite{DuAIGC, 10233667} simply feed raw prompts to the AIGC models. Due to the user's lack of experience/understanding of the specific AIGC model, outputs generated from raw prompts usually suffer from misinterpretation and limited precision \cite{YQNetwork}. Low generation quality may lead to continuous re-generation, which not only affects the Quality of Experience (QoE) but also increases the MASPs' resource consumption.
    \item \textbf{Heterogeneity of AIGC Tasks:} The current provisioning of AIGC services is static, i.e., the MASP allocates each user with equivalent computational resources for inferences and communication power to transmit outputs. However, AIGC tasks from different users exhibit significant heterogeneity. As shown in Fig. \ref{example}, even for the same task type (i.e., image generation), drawing a city with buildings is much more complex than drawing an apple on the table, since more complicated objects and compositions are involved. In this case, fixed service provisioning may lead to continuous failure of sophisticated cases, thus reducing resource efficiency.
\end{itemize}

In this paper, we present an intelligent mobile AIGC service scheme.
Specifically, to tackle the above challenges, our proposals contain interactive prompt engineering and dynamic service provisioning. 
For the first time, we integrate prompt engineering \cite{10.1145/3560815}, the cutting-edge concept to refine user prompts, into the mobile AIGC service process, with the goal of optimizing generation quality.
Additionally, we present dynamic mobile AIGC service provisioning, which trains a policy network that allows MASPs to adjust the number of inference trials and transmission power to handle each service request.
In this way, the QoE of mobile AIGC services can be significantly increased since users' requirements for high-quality AIGC outputs can be realized with lower latency and less resource consumption. 
Moreover, our scheme can be applied in any mobile AIGC application and accommodate other advanced proposals to further improve the incentive mechanism \cite{10233667} or task allocation strategy \cite{10172151}.
The contributions of this paper can be summarized as follows.
\begin{itemize}
    \item \textbf{Intelligent Mobile AIGC Services}: Different from the existing works, we reinvent the process of mobile AIGC services, evolving them for enhanced intelligence. Our goal is to maximize user QoE while reducing the resource consumption of MASPs, thus reaching the optimal system efficiency. To do so, the proposed scheme accommodates the following two mechanisms to optimize the generation quality and the service provisioning strategy.
    \item \textbf{Interactive Prompt Engineering}: To the best of our knowledge, we are the first to integrate prompt engineering into mobile AIGC services due to its well-proven ability to improve generation quality. Particularly, we address three challenges. First, the prompt should be refined based on the specific task. Hence, we leverage a Large Language Model (LLM) \cite{10.1145/3641289} to generate customized prompt corpora, with which the raw prompts can be refined precisely. Moreover, the efficacy of prompt engineering is posterior knowledge and requires substantial resources to evaluate \cite{Prompt-OIRL}. Inspired by Inverse Reinforcement Learning (IRL) \cite{zhang2023tempera}, we refine the prompt engineering policy through small-scale expert demonstrations and policy imitation. Finally, ground truth for assessing AIGC outputs might not be available due to intrinsic subjectivity. Hence, we train an LLM-based assessing agent with in-context memories to provide human-like scores for AIGC outputs and facilitate IRL training.
    \item \textbf{Dynamic Service Provisioning}: We present the problem of mobile AIGC QoE maximization, where the MASPs dynamically adjust the number of inference trials and the transmission power. Furthermore, to solve the problem, we adopt the Diffusion-Enhanced Deep Deterministic Policy Gradient (D$^3$PG) to optimize the MASP's service provisioning policy, realizing high exploration ability in varying mobile environments.
    \item \textbf{Experimental Results}: We perform extensive experiments. The numerical results demonstrate that the intelligent mobile AIGC service scheme greatly outperforms the current ones. First, prompt engineering reduces the re-generation probability by 6.3$\times$. Furthermore, dynamic service provision increases QoE by 67.8\%. The D$^3$PG also outperforms baseline algorithms in terms of reward and coverage rate.
\end{itemize}

The remainder of this paper is organized as follows.
Section II introduces the related work on mobile AIGC and discrete prompt engineering.
The system model, transmission model, and problem formulation are discussed in Section III.
Section IV demonstrates interactive prompt engineering.
Section V elaborates on the details of dynamic service provisioning via D$^3$PG.
The experiments and analysis are shown in Section VI.
Finally, Section VII concludes this paper.

\section{Related Work and Motivation}
\subsection{Mobile AIGC and Its Applications}
As a new concept, Du \textit{et al.} \cite{DuAIGC} first presented mobile AIGC and analyzed the MASP selection issues.
Then, Zhang \textit{et al.} \cite{10398474} comprehensively surveyed this topic, including its advantages, architecture, lifecycle, and some open challenges.
From 2023, mobile AIGC has entered a period of rapid development and received widespread attention from academia \cite{DuAIGC, 10233667, 10628024} and industry (e.g., Qualcomm and Meta \cite{qualcomm}).
From the model perspective, researchers keep compressing large AIGC models, reducing their costs. For instance, Qualcomm published the world's first on-device Stable Diffusion by knowledge distillation \cite{qualcomm}. Likewise, Chen \textit{et al.} \cite{chen2023speed} performed a series of GPU-aware optimizations for diffusion-based AIGC models, reducing the inference latency to three seconds. Similar proposals include LightGrad \cite{10096710}, DiffNAS \cite{li2023diffnas}, and SnapFusion \cite{li2023snapfusion}. 
To improve the efficiency of mobile AIGC networks, Xu \textit{et al.} \cite{xu2023sparks} optimized the model caching strategy of MASPs. Du \textit{et al.} \cite{10172151} presented distributed mobile AIGC inference. 
By offloading certain inference steps to users, the computation overhead of MASPs can be effectively reduced. Huang \textit{et al.} \cite{10398264} leveraged federated learning to enable mobile AIGC to generate customized content. 
Wen \textit{et al.} \cite{10233667} designed an incentive mechanism based on content freshness, thereby encouraging MASPs to reduce latency. 
Cheng \textit{et al.} \cite{AIGCSemCom} applied semantic communications to reduce the bandwidth costs of MASPs to transmit AIGC outputs.
Finally, mobile AIGC facilitates various applications. 
For example, Zhang \textit{et al.} \cite{10628024} presented a terminal-edge-cloud collaborative AIGC architecture to facilitate autonomous driving. Likewise, Zhang \textit{et al.} \cite{MATTING} designed a diffusion-based matting engine for mobile AIGC users sharing and editing content.

Different from existing works, this paper optimizes mobile AIGC from the service perspective.
By interactive prompt engineering and dynamic service provisioning, users' requests for high-quality AIGC outputs can be satisfied rapidly and consume less resources.
Hence, both the user QoE and system efficiency can be improved.

\subsection{Discrete Prompt Engineering}
Prompt engineering refers to the process of strategically refining prompts, thereby effectively guiding AIGC models to produce relevant and high-quality outputs.
According to the data structure, prompts can be split into two types, namely continuous and discrete prompts \cite{10.1145/3560815}.
The former, typically in the form of texts and images, is user-friendly and widely adopted in various AIGC applications, such as ChatGPT and Stable Diffusion.
Although the efficacy of prompt engineering in promoting generation quality has been well-proven, optimizing discrete prompts is challenging.
This is because most of the current continuous optimization approaches do not fit discrete prompt tokens.
To this end, an intuitive way is to transfer discrete prompts to continuous forms, e.g., parameterized embeddings.
Afterward, gradient-based optimization approaches can be applied \cite{wen2023hard, 10210127, ACL}.
Although improving efficiency, these methods sacrifice the interpretability of discrete prompts.
The optimized prompts cannot be explained and utilized to help users gain experience in prompting AIGC models.
Another series of proposals \cite{2309.08532, Evoprompting, deng-etal-2022-rlprompt} abstracted prompt optimization to an evaluation or Markov process.
For instance, Guo \textit{et al.} \cite{2309.08532} applied the generic algorithm, which iteratively refines each prompt by \textit{mutating} or \textit{crossing} its elements, with the goal of maximizing the fitness score.
Despite the interpretability, only limited action space and vocabulary are supported, preventing us from fully exploiting the potential of prompt engineering.

With the advancement of LLMs, refining raw prompts from infinite vocabulary becomes possible.
Hence, in this paper, we leverage an LLM to generate task-specific materials for refining raw prompts.
Moreover, to optimize the prompt engineering policy, we adopt IRL \cite{zhang2023tempera, Prompt-OIRL} to train a proxy reward.
In this way, the efficacy of selected prompt engineering strategies on any given task becomes predictable.
\renewcommand{\arraystretch}{1.2}
\begin{table}
\caption{The summary of main notations.}
\begin{tabular}{l|p{2.5cm}|l|p{2.5cm}}
\Xhline{2.2pt}
\rowcolor[rgb]{0.92,0.92,0.92}
\textbf{Notation}&\multicolumn{1}{c|}{\textbf{Description}}&\textbf{Notation}&\multicolumn{1}{c}{\textbf{Description}}\\
\hline
$Q$ & \# of users & $kc$ &Knowledge chunk\\
\hline
$M$& \# of MASPs & $\mathcal{D}$ & Demonstration dataset\\
\hline
$\pi^{(p)}_\omega$ & Prompt engineering policy& $\pi_E$ & Expert policy\\
\hline
$\pi^{(s)}_\theta$ & Service provisioning policy& $\Omega$ &AIGC model\\
\hline
$p$ & User prompt& $\tau(\cdot)$ & Embedding model\\
\hline
$\mathbf{c}_p$ & Prompt corpus& $\mathcal{D}_{\omega_1}$& Discriminator of IRL\\
\hline
$N_i$& \# of inference trials & $\mathcal{G}_\omega$& Generator of IRL\\
\hline
$P_i$& Transmission power & $\mathbf{s}^{(p)}$ & State of IRL\\
\hline
$\mathbf{p}^{*}$& Optimized prompt & $\mathbf{s}^{(s)}$ & State of D$^3$PG\\
\hline
$\otimes$& Combine operation & $T$ & \# of diffusion steps\\
\Xhline{2.2pt}
\end{tabular}
\vspace{-0.2cm}
\end{table}
\renewcommand{\arraystretch}{1}

\section{System Model}
In this section, we first introduce the intelligent mobile AIGC service scheme.
Then, the wireless transmission channel is modeled.

\subsection{General Mobile AIGC Services}
To illustrate the advantages of the proposed system, we first review a typical mobile AIGC service scheme.
Without loss of generality, this paper considers text-to-image generation, one of the most representative AIGC applications.\footnote{The proposed scheme can be extended to other AIGC applications, e.g., text-to-video, text-to-audio, and text-to-3D generation, by reformulating the prompts accordingly.}

\begin{figure}[tbp]
\centerline{\includegraphics[width=0.95\columnwidth]{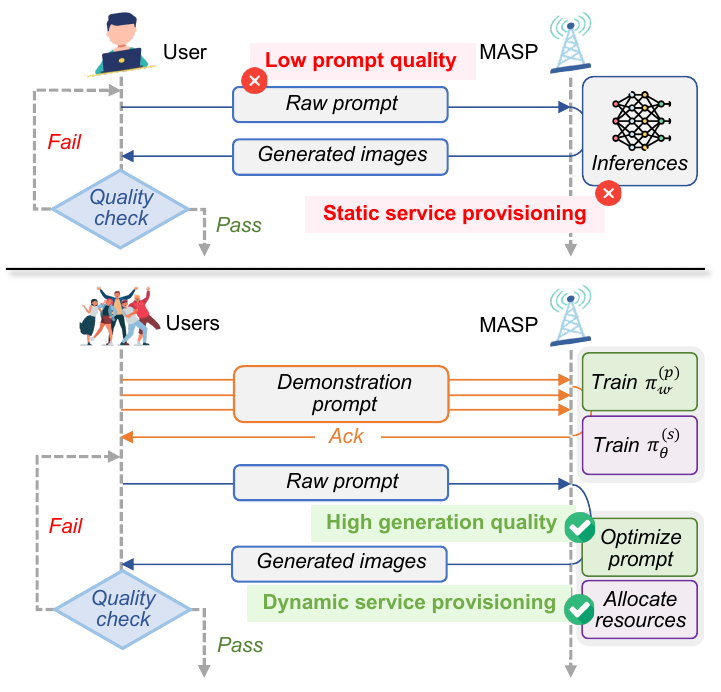}}
\caption{Top: A typical mobile AIGC service scheme (e.g., Stable Diffusion). Bottom: The proposed intelligent mobile AIGC scheme. Note that the orange and blue lines correspond to service configuration and operation stages, respectively.}
\label{structure}
\end{figure}

As illustrated in Fig. \ref{structure} (Top part), the mobile AIGC system consists of $Q$ users and $K$ MASPs, denoted as $\{U_1, \dots, U_Q\}$ and $\{M_1, \dots, M_K\}$, respectively.
To acquire AIGC images, each user first describes the required topic and style using textual prompts, which are uploaded to an MASP.
The MASP, equipped with AIGC models, performs inferences to generate a batch of images (e.g., four for Stable Diffusion\footnote{The demo is on: https://huggingface.co/spaces/stabilityai/stable-diffusion}).
Note that the users check the generation quality.
If none of the generated images reaches the users' quality requirement threshold, the MASPs will be asked to re-generate and transmit the output images again to the users.

Although this scheme can realize basic functionalities, it suffers from several issues.
Nowadays, with ever-complicated AIGC applications, directly performing inferences using raw prompts can hardly meet users' demand for pursuing high-quality and customized outputs \cite{YQNetwork}.
Frequent re-generations and re-transmissions will increase service latency and MASP's resource consumption \cite{YQNetwork}.
Moreover, the MASP allocates equal computational and communication resources for each user without considering task heterogeneity.
Thus, if complex tasks are not dynamically allocated with sufficient resources, the system efficiency will be adversely affected.
To this end, we present an intelligent mobile AIGC service scheme to improve user QoE and resource efficiency simultaneously.
\begin{figure}[tbp]
\centerline{\includegraphics[width=0.95\columnwidth]{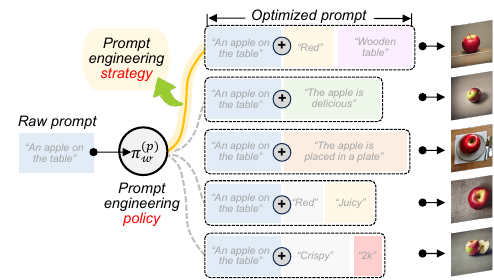}}
\caption{The illustration of prompt engineering strategy and policy. We can observe that for one raw prompt, different prompt engineering strategies lead to diverse optimized prompts and generated images. Therefore, the prompt engineering policy $\pi_\omega^{(p)}$ aims to select the optimal prompt engineering strategy dynamically.}
\label{policy}
\vspace{-0.3cm}
\end{figure}

\subsection{Intelligent Mobile AIGC Services}
As illustrated in Fig. \ref{structure} (Bottom part), our intelligent mobile AIGC consists of two stages, i.e., service configuration and service operation.

\subsubsection{Service Configuration Stage}
\begin{figure*}[tbp]
\centerline{\includegraphics[width=2\columnwidth]{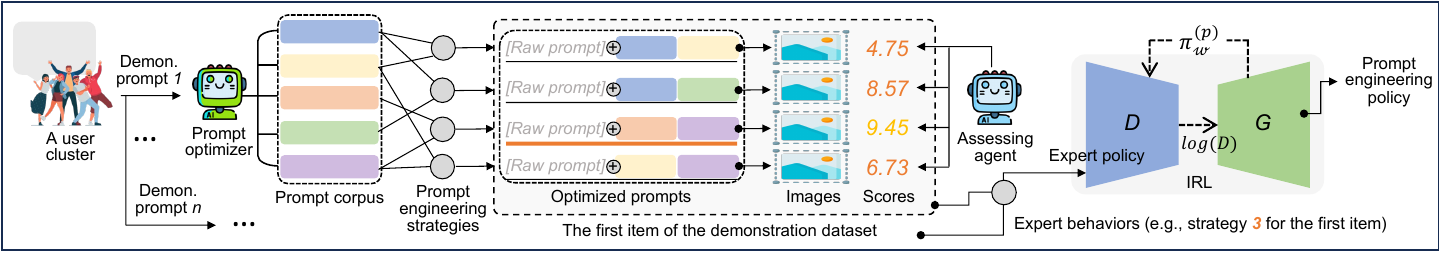}}
\caption{The workflow for training prompt engineering policy $\pi_\omega^{(p)}$. First, the prompt corpus corresponding to each demonstration prompt is generated by an LLM. Then, different prompt engineering strategies are performed, and the demonstration dataset is constructed. From the demonstration dataset, the expert policy can be acquired (The expert policy is the one that always selects the strategy that leads to the optimal generation quality). Finally, an IRL framework is utilized for policy imitation.}
\label{pipeline}
\end{figure*}

\begin{figure}[tbp]
\centerline{\includegraphics[width=0.9\columnwidth]{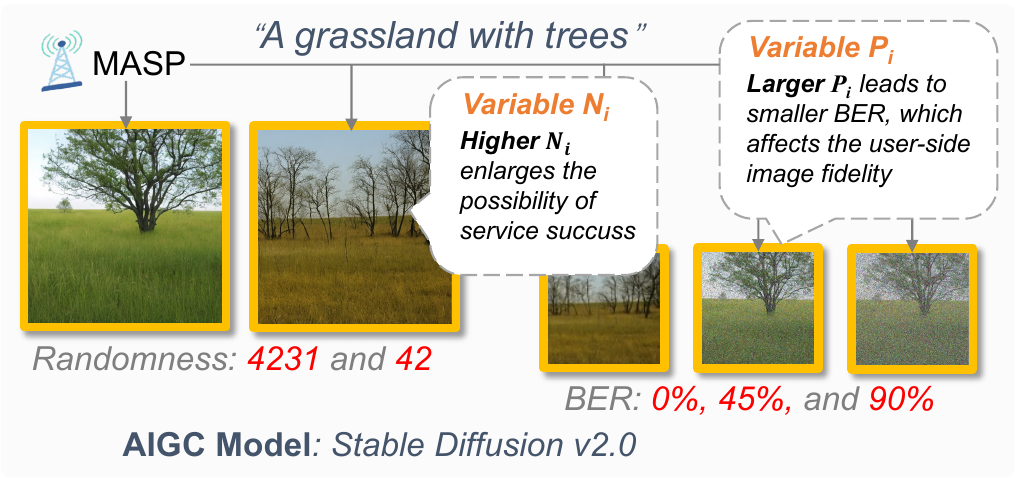}}
\caption{The impact of two decision variables of dynamic mobile AIGC service provisioning on user received images.}
\label{Variable}
\end{figure}

This stage enables the MASP to establish service policies.
First, each MASP is trained to serve a specific type of service request (e.g., \textit{generating realistic landscape photos}) \cite{9186847}.
Afterward, a customized prompt engineering policy $\pi_{\omega}^{(p)}$ optimized for this MASP should be established.
As illustrated in Fig. \ref{policy}, different prompt engineering strategies can yield varying generation qualities for the same raw prompt. 
Therefore, policy $\pi_{\omega}^{(p)}$ is designed to select the optimal prompt engineering strategy based on specific user requests and conditions, maximizing the expected generation quality. 
To effectively train $\pi_{\omega}^{(p)}$, we need to collect strategy-quality pairs that demonstrate the relationship between different actions and their outcomes.
Hence, the cluster first uploads a series of demonstration prompts to its respective MASP.
For instance, a two-item set of demonstration prompts can be [\{\texttt{A \!grassland,\! with\! trees}\}, \{\texttt{A\! lion\! sitting\! on\! a\! wooden\! bench}\}].
As shown in Fig. \ref{pipeline}, with demonstration prompts, the MASP then performs the following steps:
\begin{itemize}
    \item \textbf{Prompt Corpus Generation}: Leveraging an LLM, the MASP can generate a prompt corpus for each demonstration prompt. The corpus elements are textual segments. Then, different prompt engineering strategies can be applied, which strategically select prompt corpus elements to enrich the raw prompt.
    \item \textbf{Policy Imitation Learning}: All optimized prompts are adopted to generate images. The efficacy of all inference trials (i.e., the resulting image quality) is recorded to form a demonstrated dataset. An expert policy $\pi_{E}$ can then be acquired, which always selects the optimal strategy in the demonstration dataset (see Fig. \ref{pipeline}). Afterward, an IRL-based approach is adopted to facilitate $\pi_{\omega}^{(p)}$ imitating $\pi_{E}$, thus enabling efficient prompt engineering.
\end{itemize}
After determining the prompt engineering policy, the MASP trains another policy $\pi_\theta^{(s)}$ through D$^3$PG to dynamically provision AIGC services, with the aim of maximizing QoE.
Specifically, for each service request, $\pi_\theta^{(s)}$ solves a joint optimization problem with two decision variables, namely \textit{the number of inference trials} and \textit{the transmission power to be allocated to serve each user}, denoted as $N_{i}$ and $P_i$ ($i \in \{1, 2, \dots, Q\}$), respectively.
Fig. \ref{Variable} shows how these two factors collaborate to determine image quality on the user side.
First, the larger the number of inference trials, the higher the probability that the user acquires satisfied AIGC outputs.
The reasons are two-fold.
First, generative inference contains uncertainty and randomness.
As shown in Fig. \ref{Variable}, even using the same prompt and AIGC model, adjusting the randomness setting leads to images with totally different compositions.
Additionally, $\pi_{\omega}^{(p)}$ is an approximation to real experts rather than the optimal policy.
Hence, increasing $N_i$ can improve users' expectations of acquiring satisfying images and mitigate the effects caused by prediction errors.
Meanwhile, $P_i$ determines the Bit Error Rate (BER) of the wireless channel, which affects the fidelity of the images received by users \cite{du2023usercentric}.

\subsubsection{Service Operation Stage}
With policies $\pi_{\omega}^{(p)}$ and $\pi_\theta^{(s)}$ being trained, the MASP can provide intelligent AIGC services to mobile users.
As shown in Fig. \ref{structure} (Bottom part), for each request from $U_i$ ($i \in \{1, 2, \dots, Q\}$), the MASP first applies $\pi_{\omega}^{(p)}$ to optimize the raw prompt.
Then, dynamic service provisioning is conducted by $\pi_\theta^{(s)}$, acquiring the optimal $N_i$ and $P_i$.
$N_i$ times of generative inferences are performed, generating $N_i$ images. 
Finally, these generated images are sent to users via wireless channels using $P_i$ transmission power, accomplishing the intelligent AIGC services.

\subsection{Wireless Transmission Model}
We model the wireless transmission channel between mobile users and MASPs, considering both small-scale and large-scale fading effects \cite{9044870}. The received signal quality is influenced by fading, transmission power allocation, and channel conditions, which collectively determine the BER and image fidelity.

\subsubsection{Channel Modeling}
For small-scale fading, which results from multipath scattering, we model the channel gain using the \textit{Nakagami-m} distribution. The probability density function (PDF) of a Nakagami-\textit{m} distributed fading coefficient $X$ is given by \cite{5654629}
\begin{equation}
    f(x; \,m, \psi) = \frac{2m^m}{\Gamma(m)\,\psi^m} x^{2m-1} e^{-\frac{m}{\psi}x^2}, \quad x \geq 0,
\end{equation}
where $m$ is the fading severity parameter and $\psi = \mathbb{E}[X^2]$ is the scale parameter. The Gamma function $\Gamma(\cdot)$ is given by
\begin{equation}
    \Gamma(m) = \int_{0}^{\infty} t^{m-1} e^{-t} dt.
\end{equation}

Since the squared Nakagami-\textit{m} distributed variable $X^2$ follows a Gamma distribution, the instantaneous SNR at user $U_i$ is expressed as
\begin{equation}
    \textit{SNR}_i = \frac{P_i G_i}{N_0}.
\end{equation}
Here, $P_i$ is the allocated transmission power, $G_i = X_i^2$ represents the small-scale fading gain, and $N_0$ is the noise power. The expected SNR under Nakagami-\textit{m} fading is given by
\begin{equation}
    \mathbb{E}[\textit{SNR}_i] = \frac{P_i \psi}{N_0 }.
\end{equation}

For large-scale fading, which includes both path loss and shadowing, we model the channel gain using a log-normal distribution, i.e.,
\begin{equation}
    \boldsymbol{L}_i = d_i^{-\xi} e^{\sigma_s Z_i},
\end{equation}
where $d_i$ is the user-to-MASP distance, $\xi$ is the path-loss exponent, $\sigma_s$ is the standard deviation of the shadowing effect, and $Z_i \sim \mathcal{N}(0,1)$ is a standard normal variable representing log-normal shadowing.

Given the combined impact of small-scale and large-scale fading, the total received SNR at user $U_i$ is given by
\begin{equation}
    \textit{SNR}_i = \frac{P_i G_i}{N_0} d_i^{-\xi} e^{\sigma_s Z_i}.
\end{equation}

\subsubsection{Power Allocation and Bit Error Rate}
Given a total transmission power budget $P_{\text{total}}$ at the MASP, power is dynamically allocated among $Q$ users based on their channel conditions. The power allocated to user $U_i$ is determined as
\begin{equation}
    P_i = \frac{w_i P_{\text{total}}}{\sum_{j=1}^{Q} w_j}, \quad i \in \{1, 2, \dots, Q\},
\end{equation}
where $w_i$ is a weight factor determined by QoE requirements, channel quality, and task complexity.

The BER experienced by user $U_i$ is a function of the instantaneous SNR, which under Nakagami-\textit{m} fading \cite{5654629} is given by
\begin{equation}
    \textit{BER}_i = \int_0^\infty Q\left(\sqrt{2 \gamma}\right) f_{\textit{SNR}_i}(\gamma) d\gamma,
\end{equation}
where $Q(\cdot)$ is the standard Q-function \cite{655405}, and $f_{\textit{SNR}_i}(\gamma)$ is the probability density function of $\textit{SNR}_i$. Using the moment generating function (MGF) approach, the closed-form BER under Nakagami-\textit{m} fading can be expressed as
\begin{equation}
    \textit{BER}_i = \frac{\Gamma(m)}{2\Gamma(m+0.5)} \left( 1 - \sqrt{\frac{m}{m + \frac{\mathbb{E}[\textit{SNR}_i]}{2}}} \right)^m.
\end{equation}
Finally, the expected BER over both small-scale and large-scale fading is then computed as
\begin{equation}
    \mathbb{E}[\textit{BER}_i] = \int_{-\infty}^{\infty} \textit{BER}_i e^{-\frac{Z^2}{2}} dZ.
\end{equation}

\section{Interactive Prompt Engineering}
In this section, we detail interactive prompt engineering.
First, a prompt corpus is constructed corresponding to each demonstration prompt.
Then, we build a demonstration dataset and perform the policy imitation.
\begin{figure}[tbp]
\centerline{\includegraphics[width=0.95\columnwidth]{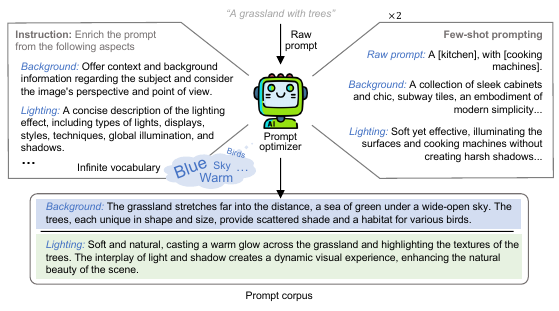}}
\caption{The prompt corpus for ``\texttt{A grassland, with trees}" considering two aspects named background and lighting. The left and right parts show the instructions and two demonstrations to $\ell_c$, respectively.}
\label{LLMprompt}
\end{figure}

\begin{figure*}[tbp]
\centerline{\includegraphics[width=1.85\columnwidth]{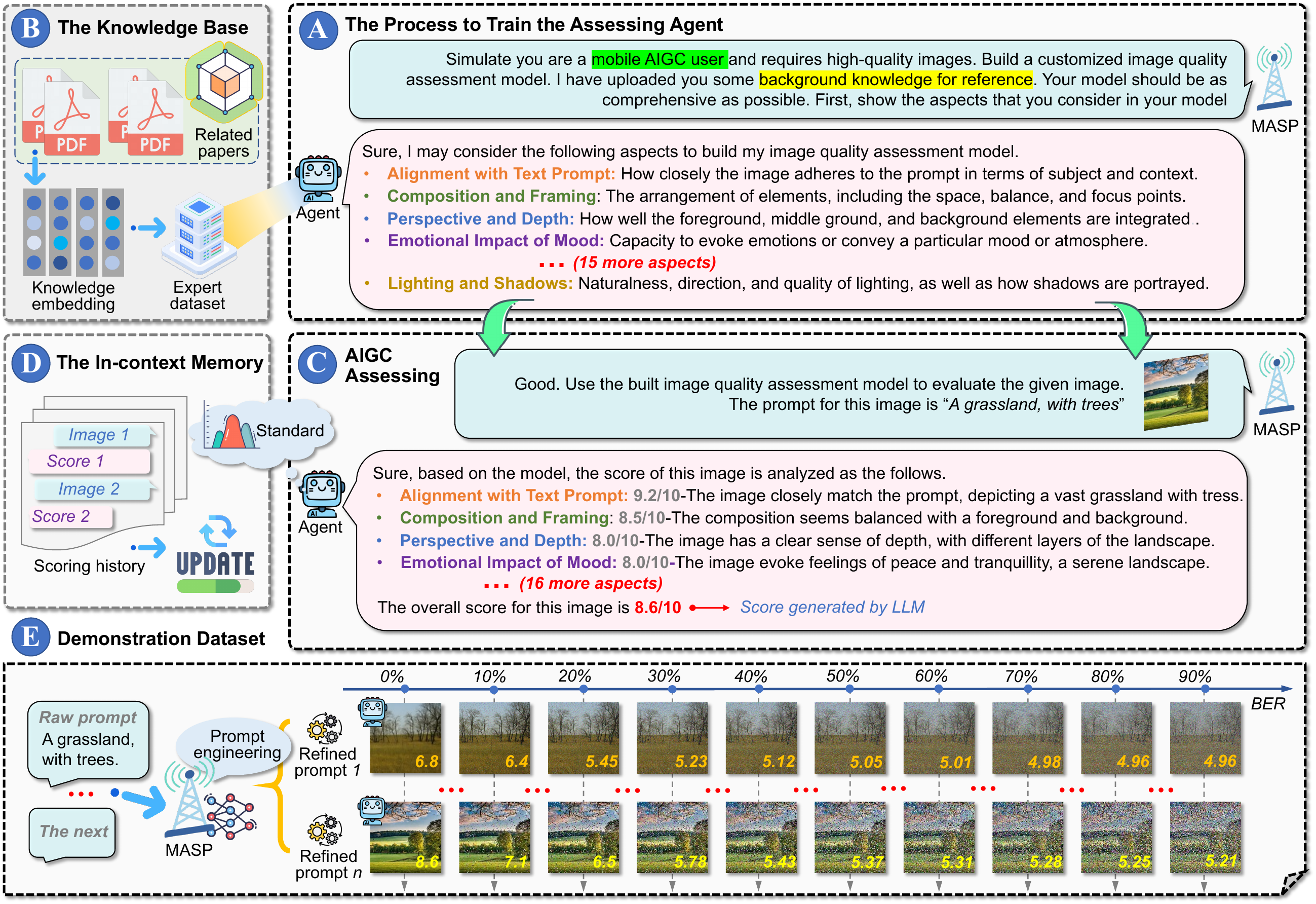}}
\caption{The LLM-based image assessment and the structure of $\mathcal{D}$. \textbf{A}: The training of the assessing agent. The prompts highlighted in green and yellow correspond to role assignment and retrieval augment, respectively. \textbf{B}: The construction of external knowledge base. \textbf{C}: The quality assessment for an image. \textbf{D}: The in-context memory. \textbf{E}: The records in $\mathcal{D}$ correspond to one demonstration prompt.}
\label{agent}
\end{figure*}

\subsection{Prompt Corpus Generation}
To support prompt engineering, the MASP will generate a $L_c$-item prompt corpus specific to each demonstration prompt $p$, denoted as $\mathbf{c}_p := \{c^{(p)}_1, c^{(p)}_2, \ldots, c^{(p)}_{L_{c}}\}$.
By decorating $p$ with materials in $\mathbf{c}_p$, more information can be fed to the text-to-image AIGC model, enabling it to retrieve more pre-learned knowledge during inferences.
Without loss of generality, we suppose that the prompts for image generations take the general form of ``\texttt{A \![a], with \![b]}", in which \texttt{[a]} and \texttt{[b]} refer to the scene and representative objects in it, respectively, e.g., ``\texttt{A [grassland], with [trees]}.\footnote{The prompt format can be freely adjusted to support different scenarios.}"
Additionally, prompt engineering follows the suffix style, i.e., appending selected elements from $\mathbf{c}_p$ as the suffix of $p$.

As shown in Fig. \ref{LLMprompt}, we leverage an LLM-based prompt optimizer $\ell_c$, such as llama2-13b-chat, to generate the prompt corpus.
Specifically, $\ell_c$ is instructed to enrich user prompts from certain aspects using infinite vocabulary, with each aspect being explained\footnote{The considered aspects are adaptable and can be customized according to the specific application and condition. The aspects considered in this paper are detailed in the Appendix. The instructions for training $\ell_c$ for enriching user prompts are published at: https://github.com/Lancelot1998/Prompt-Engineering}.
In addition, we apply two-shot prompting, i.e., feeding $\ell_c$ with two demonstrations, to regulate the required prompt corpus format.
As an example, Fig. \ref{LLMprompt} illustrates the corpus for the prompt ``\texttt{A grassland, with trees}", in which two aspects named \textit{background} and \textit{lighting} are considered.
Suppose that $k$ ($k \in \{1, 2, \dots, L_c\}$) elements are selected from $\mathbf{c}_p$ to enrich $p$.
We can derive that $\sum_{k=0}^{L_c}\left|\mathcal{P}(L_c, k)\right|$ optimized prompts can be composed by setting different arrangements of these $k$ selected elements as suffixes.
Note that $\mathcal{P}(L_c, k)$ lists the \textit{sets of permutations} on $k$ elements.
Consequently, the set of optimized prompts $\mathbf{p}^{*}$ for $L_p$ demonstration prompts $\mathbf{p} := \{p_1, p_2, \ldots, p_{L_p}\}$ can be expressed as
\begin{equation}
\mathbf{p}^{*} = \bigcup_{i=1}^{L_p} \left(\bigcup_{k=0}^{L_c} \left(\bigcup_{{\boldsymbol{\sigma}} = \mathcal{P}(L_c, k)} \left(p_i\otimes\prod_{j=1}^{k} c^{(p_i)}_{\sigma_j}\right)\right)\right),
\end{equation}
where $\sigma_j \in \boldsymbol{\sigma}$ ($j \in \{1, 2, \ldots, k\}$).
Finally, the notation $p_i\otimes\prod_{j=1}^{k}c^{(p_i)}_{\sigma_j}$ denotes the prompt engineering strategy, i.e., appending $\{c^{(p_i)}_{\sigma_1}, c^{(p_i)}_{\sigma_2}, \ldots, c^{(p_i)}_{\sigma_k}\}$ to $p_i$ as suffixes.

\subsection{Demonstration Dataset Construction}
With various candidate prompt engineering strategies, the problem becomes how to choose the best one for each request.
To optimize such a policy $\pi_{\omega}^{(p)}$, the MASP then constructs a demonstration dataset $\mathcal{D}$.
The motivation is that from the MASP's perspective, the efficacy of prompt engineering on the given prompt is a posteriori knowledge (i.e., the MASP cannot know such efficacy until it is fed back by the user) \cite{Prompt-OIRL}.
Collecting online experience during the service operation stage and polishing the prompt engineering policy from scratch is inefficient since users may suffer from low QoE during the initial time.
In contrast, constructing a demonstration dataset before formal services avoids damaging user experiences.

\subsubsection{AIGC Assessment}
Denote the AIGC model owned by MASP as $\Omega$. 
The quality assessment of the received images can be based on both quantitative metrics and user studies. 
Quantitative metrics like CLIP \cite{OpenCLIP} measure prompt-image consistency, while PicScore \cite{PicScore} evaluates aesthetic quality.
However, in real-world AIGC applications, users' quality assessments are inherently subjective, influenced by their individual perceptions, preferences, personalities, and specific requirements. 
Hence, there is no absolute ground truth for image quality assessment \cite{du2023usercentric}. 
Although user studies, e.g., questionnaires and surveys, provide subjective assessments, they present practical challenges: they are time-consuming, difficult to scale, and require repetition whenever application contexts or task patterns change.

Inspired by the recent success of LLM in agentic computing \cite{du2023usercentric}, we leverage an LLM $\ell_r$ to serve as an assessing agent, mimicking real AIGC users based on its enormous knowledge. 
Similarly to the prompt optimizer $\ell_c$, $\ell_r$ is also pluggable and can be implemented on any multimodal LLM.
As shown in Fig. \ref{agent}, we apply three techniques to train $\ell_r$, ensuring that it can give a comprehensive assessment.
\begin{itemize}
    \item \textbf{Role Prompting}: First, we train $\ell_r$ to behave like an AIGC user. Role prompting \cite{roleprompting} establishes the context and facilitates $\ell_r$ to invoke pretrained domain-specific knowledge. Hence, the generation can be aligned with the task's intent. Moreover, the specific task information is fed to $\ell_r$, including the score data structure (i.e., a floating number) and range (i.e., from 0 to 10).
    \item \textbf{Retrieval Augmentation}: In order to enrich $\ell_r$'s knowledge about image quality assessment, we build an external knowledge base with a set of documents. These include the objective factors affecting aesthetic quality, the basics of the human vision system, and the design of representative image quality assessment metrics \cite{Complexity, CLIP}. Using LangChain \cite{LangChain}, the knowledge is vectorized and divided into $W$ chunks. Hence, given the user prompt $p$, the most relevant knowledge can be fetched, i.e.,
    \begin{equation}
        p^{*} = p \otimes \!\!\!\!\!\!\underbrace{\operatorname{Top-k}}_{\text{cosine similarity}}\!\!\!\!\!\!\{kc_1, kc_2, \dots, kc_W\},
    \end{equation}
    where $kc_i$ ($i \in \{1, 2, \dots, W\}$) means a knowledge chunk. Combining pretrained and external knowledge, the assessment can be more professional.
    \item \textbf{In-Context Memory}: In real-world AIGC assessment, the user-perceivable image quality depends not only on objective and subjective factors but also on users' varying expectations according to their empirical experience. For instance, after a few service rounds, users tend to lower their expectations about difficult tasks, leading to varying levels of strictness. To reflect such a phenomenon, we equip $\ell_r$ with MemGPT \cite{MemGPT}, which saves the historical image-score pairs in the conversation memory. Then, $\ell_r$ is allowed to adjust the standard based on the context. 
\end{itemize}
With $\ell_c$ being trained, it can quantitatively assess the quality of the given image.
Furthermore, due to transmission error quantified by BER, the images received by users cannot hold 100\% fidelity.
Hence, we feed the user-received images to $\ell_r$, whose scores are called \textit{user-side score}.

\subsubsection{Data Structure}
The demonstration dataset $\mathcal{D}$ accommodates $L_p\! \cdot\! \sum_{k=0}^{L_c}\left|\mathcal{P}(L_c, k)\right|$ entries, in the form of
\begin{subequations}
\begin{flalign}
    &\mathcal{D}=\!\left\{\left[P, p_j, p^{*}_k, \mathbf{c}^{p_j}, \Upsilon\left(p^{*}_k, \Omega(p^{*}_k), P\right)\right]\right\},\\
    &j \in \{1, 2, \ldots, L_p\}, k \in \{1, 2, \ldots, |\mathbf{p}^{*}|\},
\end{flalign}
\end{subequations}
where $\Omega(p^{*}_k)$ denotes the image generated by $\Omega$ using prompt $p^{*}_k$. $\Upsilon(p^{*}_k, \Omega(p^{*}_k), P)$ represents the user-side score of $\Omega(p^{*}_k)$ transmitted using the wireless transmission power $P$, where $P \in (0, P_{\mathrm{total}}]$.
Finally, $\mathbf{p}^{*}$ has been defined in Eq. (11).

\subsubsection{Construction Process}
When constructing $\mathcal{D}$, the MASP traverses all the demonstration prompts in ${\mathbf{p}}$.
For each $p_i \in \mathbf{p}$ ($i \in \{1, 2, \ldots, L_p\}$), it applies $\ell_c$ to generate an $L_c$-element prompt corpus and perform $\sum_{k=0}^{L_c}\left|\mathcal{P}(L_c, k)\right|$ times of prompt engineering.
Afterward, the image corresponding to each optimized prompt can be generated by text-to-image model $\Omega$.
By Eq. (10), the distortion according to each possible transmission power is then applied to these images.
Finally, the user-side score for each image is assessed by $\ell_r$, and the corresponding entry is recorded in $\mathcal{D}$.

\subsection{Policy Imitation by Inverse Reinforcement Learning}
Traditionally, we can leverage $\mathcal{D}$ as an offline dataset and train $\pi^{(p)}_\omega$ using Deep Reinforcement Learning (DRL). 
Nonetheless, the actual scores are human-like subjective assessments rather than mathematically defined rewards. 
Hence, DRL can hardly effectively capture the nuanced relationships between prompt engineering strategies and generation quality from limited demonstrations. 
Instead, we leverage IRL \cite{GAIL}, which focuses on imitating expert policies by learning from expert behaviors, enabling us to better capture the subjective nature of AIGC quality assessment while maximizing sample usage efficiency \cite{Prompt-OIRL}.
Following the IRL principle, we first define the state and action spaces of our task.
\begin{itemize}
    \item \textbf{States}: The state describes the environment with which the prompt engineering policy interacts. Let $\mathbf{s}^{(p)}_{t}$ denote the IRL state at moment $t$, it can be expressed as 
        \begin{equation}
            \mathbf{s}_t^{(p)} = \left\{ \textbf{h}, \tau(p_i),  P_i \right\},
        \end{equation}
    where $\textbf{h} = \{a_1^{(p)}, a_2^{(p)}, \dots, a_{t-1}^{(p)}\}$ is the history of actions taken from genesis moment to moment $t\!-\!1$. $\tau(\cdot)$ refers to the embedding function \cite{du2023usercentric}, which converts a natural language prompt into machine-friendly vectors. $P_i$ is the allocated transmission power that affects BER.
    \item \textbf{Action}: The action space consists of all available prompt engineering strategies to refine the given raw prompt, which can be defined as
        \begin{equation}
            \mathbf{a}_t^{(p)}({p}_i) = \mathbb{S}\left(\bigcup_{k=0}^{L_c} \left(\bigcup_{{\boldsymbol{\sigma}} = \mathcal{P}(L_c, k)} \left(p_i\otimes\prod_{j=1}^{k} c^{(p_i)}_{\sigma_j}\right)\right)\right).
        \end{equation}
    Note that $\mathbb{S}(\cdot)$ represents an empirical filter. Note that we adopt $\mathbb{S}(\cdot)$ because given the large combinations of prompt corpus elements, in practice, we only consider the most representative prompt engineering strategies (the details are discussed in Section VI).
\end{itemize}
As aforementioned, the reward $\mathcal{R}(\mathbf{a}, \mathbf{s})$ in our problem is unknown.
Nonetheless, based on the demonstration dataset $\mathcal{D}$, an expert prompt engineering policy $\pi_E$ can be established, which maximizes the actual reward of all demonstration prompts by selecting the best strategy.
$\pi_E$ can be expressed as
\begin{equation}
    \max_{p^{*}_k} \Upsilon\left(p^{*}_k, \Omega(p^{*}_k\right), P^{(i)}), \;\;\forall k \in \{1, 2, \dots, |\mathbf{p^{*}}|\}.
\end{equation}
We optimize $\pi_\omega^{(p)}$ by letting it imitate $\pi_E$.
To do so, inspired by Generative Adversarial Imitation Learning (GAIL) \cite{GAIL}, we construct a generator-discriminator architecture to optimize $\pi_\omega^{(p)}$ adversarially.
Specifically, the discriminator $\mathcal{D}_{\omega_1}$ is a bi-classifier that distinguishes the actions sampled from policies $\pi_E$ and $\pi_\omega^{(p)}$.
Consequently, the objective function of $\mathcal{D}_{\omega_1}$ can be defined as
\begin{equation}
    \max_{\mathcal{D}_{\omega_1}} \mathbb{E}_{\pi_E}\!\!\left[\log \mathcal{D}_{\omega_1}(\mathbf{s}^{(p)}, \mathbf{a}^{(p)})] \!+\! \mathbb{E}_{\pi_{\omega}^{(p)}}[\log (1 \!-\! \mathcal{D}_{\omega_1}(\mathbf{s}^{(p)}, \mathbf{a}^{(p)}))\right],
\end{equation}
where $\mathcal{D}_{\omega_1}(\mathbf{s}^{(p)}, \mathbf{a}^{(p)}) \in \{0, 1\}$.

The generator $\mathcal{G}_\omega$ aims to refine $\pi_\omega^{(p)}$ towards imitating $\pi_E$.
Hence, the cost function can be defined as $\mathbb{E}_{\pi^{(p)}_\omega}[\log (1 - \mathcal{D}_{\omega_1}(\mathbf{s}^{(p)}, \mathbf{a}^{(p)}))]$ i.e., minimizing the success rate of $\mathcal{D}_{\omega_1}$.
Then, we leverage Proximal Policy Optimization (PPO) \cite{10032267} as the policy optimization framework due to its stability in learning. 
PPO evaluates the efficiency of the current policy via an advantage function, which is defined as
\begin{equation}
    \hat{\mathcal{A}}(\mathbf{s}^{(p)}, \mathbf{a}^{(p)}) = r_t + \gamma \mathcal{V}_{\phi}(\mathbf{s}^{(p)}_{t+1}) - \mathcal{V}_{\phi}(\mathbf{s}^{(p)}_t),
\end{equation}
where $r_t$ refers to the direct reward of the current policy, i.e., $\mathbb{E}_{\pi_{\omega}^{(p)}}[\log (1 - \mathcal{D}_{\omega_1}(\mathbf{s}^{(p)}, \mathbf{a}^{(p)}))]$.
$\gamma$ represents the discount factor for future rewards.
$\mathcal{V}_{\phi}(\cdot)$ means the state value function predicted by the PPO critic network.
Then, the objective function can be defined as
\begin{equation}
    \mathcal{L}^{CLIP}(\omega) = \mathbb{E}_t\left[\min(r_t(\omega)\hat{\mathcal{A}}_t, \text{clip}(r_t(\omega), 1-\epsilon, 1+\epsilon)\hat{\mathcal{A}}_t)\right],
\end{equation}
where $r_t(\omega) = \frac{\pi_\omega^{(p)}(\mathbf{a}^{(p)}_t|\mathbf{s}^{(p)}_t)}{\pi_{\omega_{old}}^{(p)}(\mathbf{a}^{(p)}_t|\mathbf{s}^{(p)}_t)}$, referring to the probability ratio between the current policy $\pi_\omega$ and the old policy $\pi_{\omega_{old}}$.
$\epsilon$ represents the clipping parameter that bounds policy updates to prevent excessive changes.
Note that the clip($\cdot,\cdot,\cdot$) function \cite{10032267} ensures that the objective function remains within a reasonable range by limiting the probability ratio between $[1-\epsilon, 1+\epsilon]$, which stabilizes training and prevents destructive policy changes that could deviate significantly from expert behaviors.
With $\mathcal{L}^{CLIP}$, the generator $\mathcal{G}_\omega$ can be updated by 
\begin{equation}
    \omega' = \omega + \alpha \nabla_\omega\mathcal{L}^{CLIP}(\omega),
\end{equation}
where $\alpha$ means the learning rate for updating the generator network.
Finally, the critic network is updated by
\begin{equation}
\phi' = \phi - \beta\nabla_\phi\mathbb{E}_t\left[(\mathcal{V}_\phi(\mathbf{s}^{(p)}_t) - (r_t + \gamma \mathcal{V}_\phi(\mathbf{s}^{(p)}_{t+1})))^2\right],
\end{equation}
where $\beta$ is the learning rate for the critic network, and $\gamma$ is the discount factor for future rewards.

\section{Diffusion-Empowered Dynamic Service Provisioning}
In this section, we detail the proposed dynamic service provisioning.
First, we formulate the problem and model the QoE of mobile AIGC users.
Then, we proposed the D$^3$PG to generate the optimal service provisioning policy.

\subsection{Problem Formulation}
The MASP aims to achieve an optimal balance between user QoE and resource efficiency, including computing resource allocation to perform prompt engineering and transmission power to transmit AIGC outputs. 
This problem can be formulated as follows:
\begin{subequations}
\begin{flalign}
\max _{\{N_{i}, P_{i}\}} & \sum_{i=1}^{Q} \left(\eta_q\cdot\mathcal{Q}\left(N_i, P_i\right) - \eta_c\cdot\mathcal{C}\left(N_i, P_i\right)\right), \\
\text { s.t., } & \mathcal{Q}\left(N_i, P_i\right) \geq \mathcal{Q}^{\mathrm{th}}_i, \quad \forall i \in\{1, 2, \ldots, Q\}, \\
& N_{i}\, \geq 1, \quad \forall i \in\{1, 2, \ldots, Q\},\\
& \sum_{i=1}^{Q} P_i \leq P_{\mathrm{total}},
\end{flalign}
\end{subequations}
where $\mathcal{Q}(\cdot)$ and $\mathcal{C}(\cdot)$ denote the functions for QoE and cost calculation, respectively.
$\eta_q$ and $\eta_c$ are two weighting factors.
The constraint in Eq. (22b) indicates that the QoE of each user should meet its requirement threshold.
The constraint in Eq. (22c) defines the range of $N_i$, i.e., the MASP should generate at least one image each time.
Finally, Eq. (22d) requires that the total transmission power allocated by the MASP cannot exceed its budget.
In the following parts, we elaborate on the modeling of $\mathcal{Q}(\cdot)$ and $\mathcal{C}(\cdot)$, respectively.

\subsection{QoE Modeling}
In mobile AIGC, the user QoE mainly depends on two key performance indicators, namely service latency and generation quality.
The former is related to the number of inference trials and the time required for each round of inference.
The latter, as mentioned in Section III, is determined by the efficacy of prompt engineering and the transmission power that affects the fidelity of the user's received images.
Jointly considering the above factors, the QoE for user $U_i$ is defined as
\begin{equation} 
    \mathcal{Q}(N_{i}, P_i) \!=\! \overbrace{\operatorname{log}_{N_i}\!\!\underbrace{\left(\frac{L_{max}}{N_i\cdot T_\zeta}\right)}_{\textnormal{service latency}}}^{\textnormal{impact of latency on QoE}}\operatorname{ln}\left(\frac{\max \left\{Q^{(i)}_1, \dots, Q^{(i)}_{N_i}\right\}}{Q_\mathrm{th}^{(i)}}\right), 
\end{equation}
where $T_\zeta$ represents the inference time with $\zeta$ denoting the number of diffusion steps.
$L_\mathrm{max}$ and $Q^{(i)}_\mathrm{th}$ denote the upper bound of service latency and $U_i$'s personal threshold for generation quality, respectively.
Suppose that the user only adopts the most satisfied AIGC output.
Hence, we apply a filter and fetch the maximum generation quality from $\{Q^{(i)}_1, \dots, Q^{(i)}_{N_i}\}$.
Note that we leverage the method in \cite{6263849} to model users' tolerance towards service latency.
Specifically, $N_i\cdot T_\zeta$ means the total inference time for $N_i$ trials\footnote{For simplicity, we ignore the transmission latency and suppose service latency equals inference time since it is the major latency cause.}.
In \cite{6263849}, Hossfeld \textit{et al.} proved that the subjective impact of service latency on user experience follows a $\operatorname{log}$ relationship, i.e., as the waiting time increases, users will become less sensitive towards latency increment.
Moreover, the larger $N_i$ is, the higher the user's tolerance for service latency since more images can be received in this round.
Therefore, we apply $N_i$ as the base of the logarithmic function.

Moreover, to effectively model the user's subjective experience toward generation quality, we apply Weber-Fechner law \cite{WF_Law}.
This law states that as the stimulus (e.g., the vision, hearing, taste, and touch) increases, the perceived sensation grows but at a diminishing rate. 
Similar to \cite{6263849}, such a phenomenon is described as a logarithmic relationship.
In addition, the noticeable difference between two different levels of stimuli is a constant ratio of the initial stimulus.
To this end, we define the impact of generation quality on overall QoE as $\operatorname{ln}\left(\frac{\max \left\{Q^{(i)}_1, \dots, Q^{(i)}_{N_i}\right\}}{Q_\mathrm{th}^{(i)}}\right)$, as illustrated in Eq. (23).

Up till now, we have defined the QoE function $\mathcal{Q}(N_i, P_i)$.
Another consideration of system efficiency is the resource consumption of the MASPs, containing the computation resources to perform generative inference and the transmission power to transmit generated images to users.
Hence, $\mathcal{C}(N_i, P_i)$ can be defined as
\begin{equation}
    \mathcal{C}(N_i, P_i) = N_i \cdot \left(c_\zeta + P_i\right),
\end{equation}
where $c_\zeta$ represents the computation resource consumption for each generative inference trail, with $\zeta$ meaning the diffusion step number.
Substituting Eqs. (23) and (24) into Eq. (22a), we can obtain the complete objective about joint QoE and resource optimization.
Next, we design a diffusion-based approach to generate the optimal solution to this problem.

\subsection{Algorithm Overview}
The proposed D$^3$PG follows a DRL architecture with five basic components, namely agent, state, action, policy, and reward.
Their introductions are shown below.
\begin{itemize}
    \item \textbf{Agent}: Our agent is the MASP, which performs the service provisioning to allocate the physical resources to serve $Q$ mobile users simultaneously.
    \item \textbf{State}: The state of the mobile AIGC environment takes the form of $\mathbf{s}^{(s)}$ := [\{$\tau(p_1)$, $\tau(p_2)$, $\dots$, $\tau(p_Q)$\}, \{$d_1$, $d_2$, $\dots$, $d_Q$\}, \{$\mathcal{Q}^{\mathrm{th}}_1$, $\mathcal{Q}^{\mathrm{th}}_2$, $\dots$, $\mathcal{Q}^{\mathrm{th}}_Q$\}, $P_{\mathrm{total}}, \;\widetilde{\textit{SNR}}$]. The first two sets accommodate the prompts and distances from the MASP to users $\{U_1, U_2, \dots, U_Q\}$, respectively. $P_{\mathrm{total}}$ represents the MASP's total transmission power, and $\widetilde{\textit{SNR}}$ is the wireless channel state, as explained in Section III.
    \item \textbf{Action}: We define the action space as a vector $\mathbf{a}^{(s)} := \{\boldsymbol{a}^{(s)}_1, \boldsymbol{a}^{(s)}_2, \dots, \boldsymbol{a}^{(s)}_Q\}$, denoting the resources allocated to each user. Specifically, each $\boldsymbol{a}^{(s)}_i := \{N_i, P_i\}$ ($\forall i \in \{1, 2, \dots, Q\}$), including the number of inference trials and the allocation transmission power.
    \item \textbf{Policy}: The policy refers to the probability that the agent takes action $\mathbf{a}^{(s)}$ in the state $\mathbf{s}^{(s)}$. Particularly, our algorithm adopts a diffusion network parameterized by $\theta$ to learn the relationship between the input state $\mathbf{s}^{(s)}$ and the output action $\mathbf{a}^{(s)}$ that can optimize the reward. Therefore, this policy network can be expressed as $\pi^{(s)}_\theta(\mathbf{s}^{(s)}, \mathbf{a}^{(s)}) = \operatorname{Prob}(\mathbf{a}^{(s)}|\mathbf{s}^{(s)})$.
    \item \textbf{Reward}: Finally, given the state space $\mathbf{s}^{(s)}$, the reward of taking action $\mathbf{a}^{(s)}$ can be defined as $R(\mathbf{a}^{(s)}|\mathbf{s}^{(s)}) = \sum_{i=1}^{Q} \left(\eta_q\cdot\mathcal{Q}\left(N_i, P_i\right) - \eta_c\cdot\mathcal{C}\left(N_i, P_i\right)\right)$, i.e., Eq. (22a). Note that if any of the constraints shown in Eqs. (22b)-(22d) is not satisfied, we apply a negative penalty. Specifically, if the actions for $J$ users fail to meet Eqs. (22b) or (22c), the penalty is $J \cdot \varrho$, where $\varrho$ is a hyperparatermer. If Eq. (20d) is not satisfied, the penalty becomes $Q \cdot \varrho$ because the generated service provisioning solution makes the problem infeasible. 
\end{itemize}

\subsection{Diffusion-Enhanced DDPG (D$^3$PG) Design}
\subsubsection{Diffusion-Empowered Policy Generation}
Inspired by non-equilibrium thermodynamics, diffusion models characterize the generation tasks as a step-by-step process of denoising from pure Gaussian noise \cite{DiffusionDRL1}.
Nowadays, diffusion has supported numerous AIGC models in various modalities, such as the Stable Diffusion we used in Fig. \ref{example}.
Additionally, it brings traditional DRL algorithms with greater exploration ability \cite{DiffusionDRL1, zhu2023diffusion}.
Therefore, our D$^3$PG employs a deep diffusion network to generate policy $\pi^{(s)}_\theta(\mathbf{s}^{(s)}, \mathbf{a}^{(s)})$.
Specifically, the network contains two Markov processes, namely forward diffusion and denoising.
The former perturbs the optimal action $\mathbf{a}^{(s)}_0$ to random action $\mathbf{a}_T$ by $T$ diffusion steps, satisfying
\begin{equation}
    \mathbf{a}^{(s)}_t = \sqrt{\alpha_t}\mathbf{a}^{(s)}_{t-1} + \sqrt{1-\alpha_t}\epsilon_t, \quad \epsilon \sim \mathcal{N}(0, \mathbf{I}),
\end{equation}
where $\mathbf{I}$ denotes the identity matrix. $\alpha_t$ ($t \in \{1, 2, \dots, T\}$) follows a pre-defined schedule and is decreasing over $t$ \cite{DDPM}.
Hence, the entire forward diffusion can be expressed as
\begin{subequations}
\begin{flalign}
    q(\mathbf{a}^{(s)}_{1:T}|&\mathbf{a}^{(s)}_0) = \prod_{t=1}^{T} q(\mathbf{a}^{(s)}_t|\mathbf{a}^{(s)}_{t-1}), \\
    q(\mathbf{a}^{(s)}_t|\mathbf{a}^{(s)}_{t-1}) =& \mathcal{N}(\mathbf{a}^{(s)}_t; \sqrt{\alpha_t}\mathbf{a}^{(s)}_{t-1}, (1-\alpha_t)\,\mathbf{I}).
\end{flalign}    
\end{subequations}
Accordingly, the denoising process, i.e., generating the optimal policy from noise, can be expressed as \cite{DDPM}
\begin{equation}
    \begin{split}
        p_\theta(\mathbf{a}^{(s)}_{0:T}) &= p(\mathbf{a}^{(s)}_T)\prod^{T}_{t=1}p_\theta(\mathbf{a}^{(s)}_{t-1}|\mathbf{a}^{(s)}_t). \\
    \end{split}
\end{equation}
Such a process can be trained by maximizing the likelihood of $p_\theta(\mathbf{a}_{0})$.
However, $p_\theta(\mathbf{a}^{(s)}_{t-1}|\mathbf{a}^{(s)}_t)$ cannot be directly calculated.
To this end, $q(\mathbf{a}^{(s)}_{t-1}|\mathbf{a}^{(s)}_t, \mathbf{a}^{(s)}_0)$ is employed.
Suppose that $q(\mathbf{a}^{(s)}_{t-1}|\mathbf{a}^{(s)}_t, \mathbf{a}^{(s)}_0)$ follows the normal distribution.
Applying the Bayesian formula and Eq. (26), the mean and variance can be calculated as \cite{DDPM}
\begin{subequations}
\begin{flalign}
    q(\mathbf{a}^{(s)}_{t-1}|\mathbf{a}^{(s)}_t, \mathbf{a}^{(s)}_0)  &\!=\! \mathcal{N}\!\left(\mathbf{a}^{(s)}_{t-1}; \mu_t(\mathbf{a}^{(s)}_t, t), \Sigma_t(\mathbf{a}^{(s)}_t, t)\right),\\
    \mu_t(\mathbf{a}^{(s)}_t, t) &= \frac{1}{\sqrt{\alpha_t}}\left(\mathbf{a}^{(s)}_t - \frac{1-\alpha_t}{\sqrt{1-\bar{\alpha}_t}}\epsilon\right),\\
    \Sigma_t(\mathbf{a}^{(s)}_t, t) &= \frac{(1-\alpha_t)(1-\bar{\alpha}_{t-1})}{1-\bar{\alpha}_t} \,\mathbf{I},
\end{flalign}
\end{subequations}
where $\bar{\alpha}$ = $\prod_{s=1}^{t}\alpha_s$
With $q(\mathbf{a}^{(s)}_{t-1}|\mathbf{a}^{(s)}_t, \mathbf{a}^{(s)}_0)$, the variational lower bound of $\operatorname{log}p_\theta(\mathbf{a}^{(s)}_0)$ can be calculated.
The final training objective can be derived as 
\begin{equation}
    \min ||\epsilon - \epsilon_\theta(\sqrt{\bar{\alpha}_t}\mathbf{a}^{(s)}_0 + \sqrt{1-\bar{\alpha}_t}\epsilon, t)||^2,
\end{equation}
where $\epsilon_\theta$ contains the parameters (implemented by a UNet) to be trained \cite{DDPM}.
After training, the optimal action $\mathbf{a}^{(s)}_0$ can be generated step-by-step from a random one $\mathbf{a}^{(s)}_T$, i.e.,
\begin{equation}
    \mathbf{a}^{(s)}_{t-1} = \frac{1}{\sqrt{\alpha_t}}\left(\mathbf{a}^{(s)}_t - \frac{1-\alpha_t}{\sqrt{1-\bar{\alpha}}_t}\epsilon_\theta(\mathbf{a}^{(s)}_t, t)\right),
\end{equation}
where $t \in \{1, 2, \dots, T\}$.
\begin{figure}[tbp]
\centerline{\includegraphics[width=0.9\columnwidth]{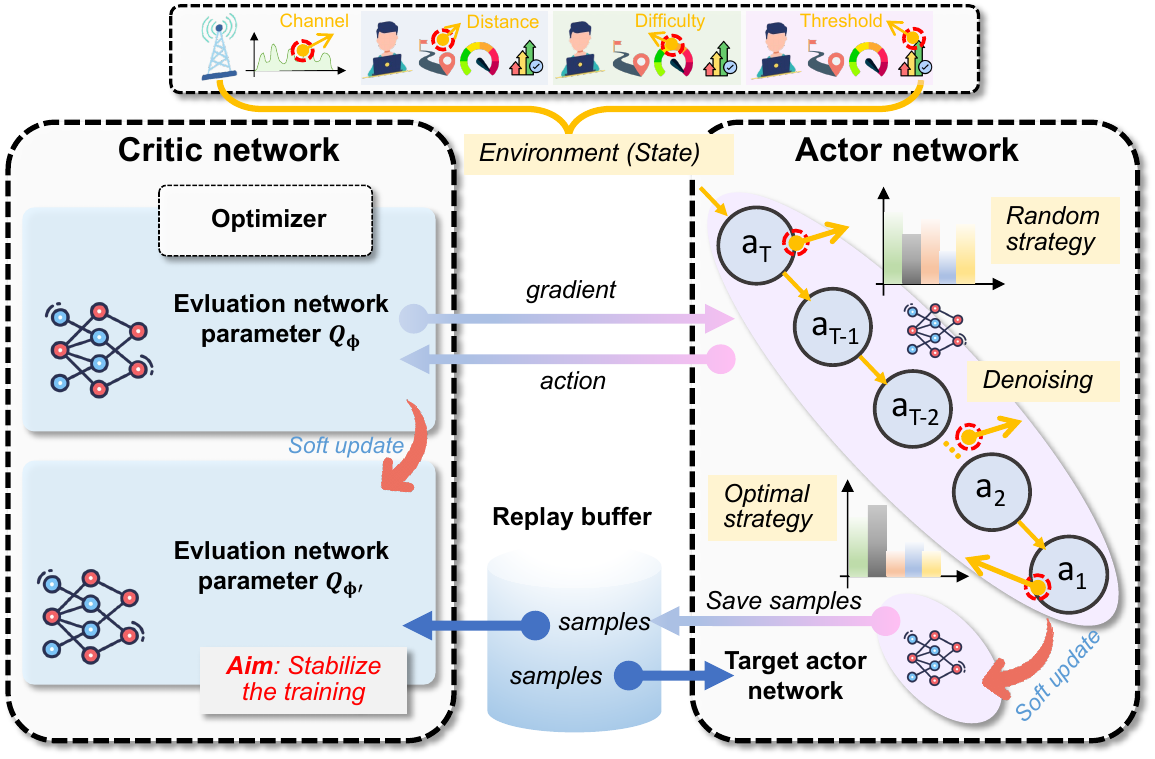}}
\caption{The D$^3$PG architecture. We apply a diffusion-based actor-network to enhance the DDPG.}
\label{ddpm}
\end{figure}

\subsubsection{Model Architecture}
We utilize the DDPG \cite{DiffusionDRL1} architecture to accommodate the diffusion-based policy network, forming D$^3$PG.
As shown in Fig. \ref{ddpm}, diffusion acts as the actor networks, which generate service provisioning strategies and interact with the mobile AIGC environments.
In addition, two critic networks are employed, using the Bellman equation to estimate the expected reward, i.e.,
\begin{equation}
    Q_\phi(\mathbf{s}^{(s)}\!, \mathbf{a}^{(s)}) =\! R(\mathbf{a}^{(s)}|\mathbf{s}^{(s)}) + \gamma Q_\phi'\left(\mathbf{s}^{(s)'}, \pi^{(s)}_{\theta'}(\mathbf{s}^{(s)'})\right),
\end{equation}
where $\mathbf{s}^{(s)'}$ denotes the next state, $\phi'$ and $\theta'$ represent the parameters of the target networks for actor and critic, respectively, and $\gamma$ is the discount factor.
Note that in DDPG, target networks for both the actor and the critic are applied to stabilize the training process. 
These target networks have the same architecture as the original networks, but their weights are updated slowly, usually by soft updates.
The policy update aims to maximize the Q-value, which can be expressed by
\begin{equation}
    \max_{\pi_\theta} \mathbb{E}_{\mathbf{a}^{(s)}\sim\pi_{\theta}}\left[Q_{\pi}(\mathbf{s}^{(s)}, \mathbf{a}^{(s)})\right].
\end{equation}
The detailed training process is shown in \textbf{Algorithm 1}.
\begin{algorithm}[tpb]
\footnotesize \caption{The Procedure of D$^3$PG Algorithm}
\begin{algorithmic}[1]
\Require  
$\mathbf{s}^{(s)}$, $N_b$, $T$, $\eta$, $\gamma$ \textit{\#\#\, The mobile AIGC environment, batch size, diffusion step number, discount factor, and learning rate}
\Ensure 
$\mathbf{a}_0$ \textit{\#\#\, service provisioning strategy}
\Procedure{Algorithm Training}{$\mathbf{s}^{(s)}$, $N_b$, $T$, $\eta$, $\gamma$} 
\State Initialize networks: actor network $\pi^{(s)}_{\theta}$ and critic networks $\phi$ and $\phi$'.
\While{not converged}
\State Initialize random noise $\mathbf{a}^{(s)}_T$; generate bandwidth allocation scheme $\mathbf{a}^{(s)}_0$ by denoising process shown in Eq. (30).
\State Add exploration noise to $\mathbf{a}^{(s)}_0$.
\State Execute service provisioning and calculate reward $R(\mathbf{a}^{(s)}|\mathbf{s}^{(s)})$ by Eq. (22a).
\State Store the record ($\mathbf{s}^{(s)}, \mathbf{a}_0^{(s)}, R(\mathbf{a}^{(s)}|\mathbf{s}^{(s)})$) in the replay buffer
\State Randomly select $N_b$ records
\State Update the policy generation network
\State Update the Q-networks
\EndWhile
\EndProcedure
\Statex
\Procedure{Algorithm Inference}{$\mathbf{s}^{(s)}$, $N_b$, $T$, $\eta$, $\gamma$}
\State Observe the environment $\mathbf{s}^{(s)}$
\State Generate bandwidth allocation scheme $\mathbf{a}^{(s)}_0$
\State \textbf{Return} $\mathbf{a}^{(s)}_0$
\EndProcedure
\end{algorithmic}
\end{algorithm}

\subsubsection{Complexity Analysis}
We then examine the computational complexity of D$^3$PG in detail. 
Frist, suppose that $S_p$ and $S_q$ respectively represent the sizes of the diffusion-based actor-network and the Q-network. 
The architectural complexity is $\mathcal{O}(S_p + 2S_q)$. 
Because each service provisioning solution should be generated through $T$ rounds of diffusion denoising, the complexity of generating each action is $\mathcal{O}(T S_p)$. 
Consequently, the overall complexity is $\mathcal{O}((T + 1) S_p + 2S_q)$. Furthermore, if $\delta$ training epochs are performed with a batch size of $S_b$, the resulting computational cost is $\mathcal{O}(\delta S_b \bigl((T + 1) S_p + 2S_q\bigr))$. Finally, during the inference stage, the complexity amounts to $\mathcal{O}(S_p)$.


\renewcommand{\arraystretch}{1.2}
\begin{table}[pb]
\caption{The involved prompt engineering strategies.}
\begin{tabular}{p{2.3cm}|p{5.5cm}}
\Xhline{2.2pt}
\rowcolor[rgb]{0.92,0.92,0.92}
\textbf{Strategy} & \multicolumn{1}{c}{\textbf{Description}} \\
\hline
\multirow{1}{*}{Strategy 0} & \textit{Raw prompt} \\
\hline
\multirow{1}{*}{Strategy 1} & \textit{Object description} \\
\hline
\multirow{1}{*}{Strategy 2} & \textit{Object description + environment} \\
\hline
\multirow{1}{*}{Strategy 3} & \textit{Object description + mood} \\
\hline
\multirow{1}{*}{Strategy 4} & \textit{Object description + lighting} \\
\hline
\multirow{1}{*}{Strategy 5} & \textit{Object description + quality booster} \\
\hline
\multirow{1}{*}{Strategy 6} & \textit{Object description + negative effects} \\
\Xhline{2.2pt}
\end{tabular}
\end{table}
\renewcommand{\arraystretch}{1}
\renewcommand{\arraystretch}{1.2}
\begin{table}[htpb]
\caption{The experimental settings.}
\begin{tabular}{p{1.5cm}|p{2.3cm}|p{3.6cm}}
\Xhline{2.2pt}
\rowcolor[rgb]{0.92,0.92,0.92}
\textbf{Parameter} & \textbf{Description} & \textbf{Value} \\
\hline
\multirow{1}{*}{$\ell_c$} &Prompt optimizer& \textit{ChatGPT (GPT-3.5-turbo)} \\
\hline
\multirow{1}{*}{$\ell_r$} &Assessing agent& \textit{GPT-4-vision-preview} \\
\hline
\multirow{1}{*}{$\Omega$} &AIGC model& \textit{Stable Diffusion v2.0} \\
\hline
\multirow{1}{*}{$\zeta$} &Diffusion step& \textit{25} \\
\hline
\multirow{1}{*}{$Q$} &\# of users& \textit{3} \\
\hline
\multirow{1}{*}{$M$} &\# of MASP& 1 \\
\Xhline{2.2pt}
\end{tabular}
\vspace{-0.5cm}
\end{table}
\renewcommand{\arraystretch}{1}
\begin{figure*}[tbp]
\centerline{\includegraphics[width=1.95\columnwidth]{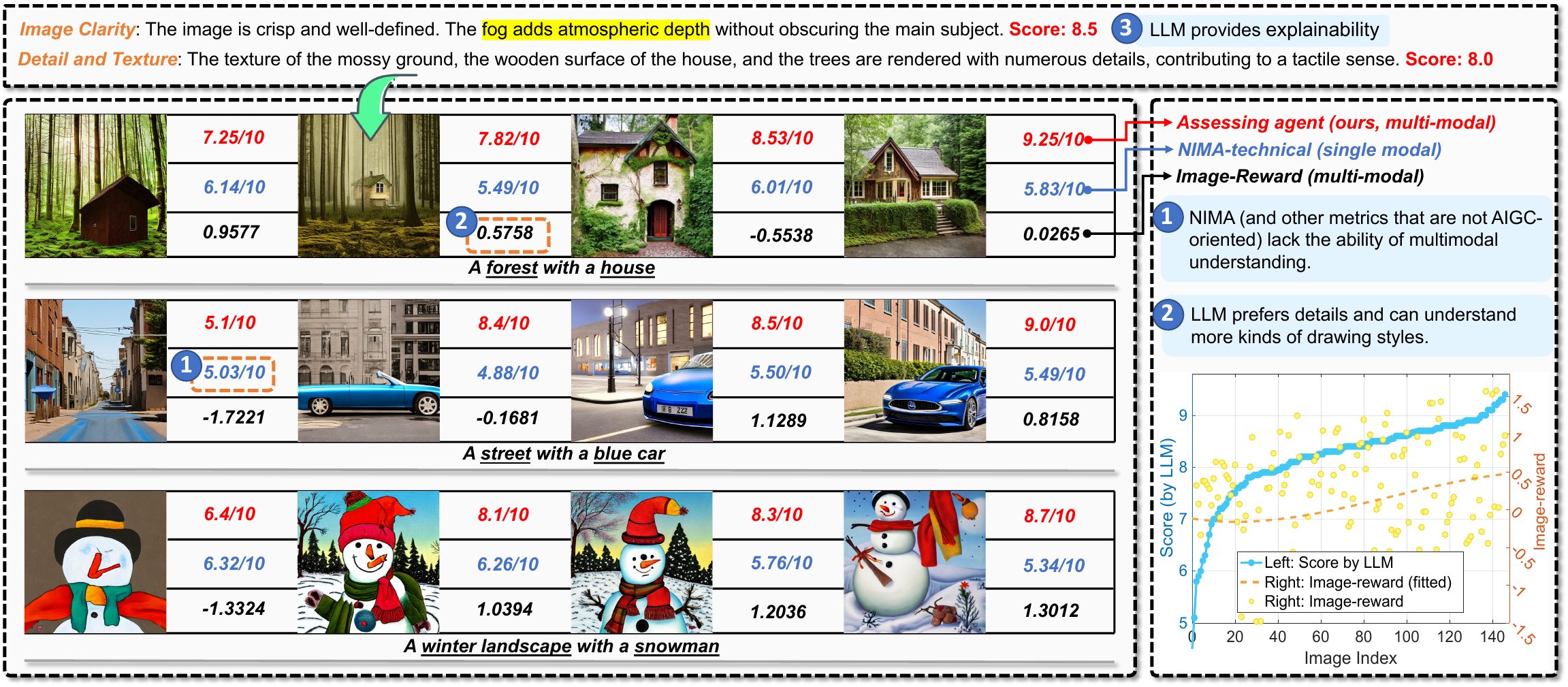}}
\caption{The rationale of LLM-empowered assessing agent. Red, blue, and black scores are from the assessing agent (ours), NIMA, and image-reward, respectively. Note that the images in the same row are sorted in ascending order of the assessing agent's score.}
\label{rationale}
\vspace{-0.3cm}
\end{figure*}

\section{Performance Evaluation}

\textbf{Testbed.} The experiments are conducted on a server with three NVIDIA RTX A5000 GPUs with 24 GB of memory and an AMD Ryzen Threadripper PRO 3975WX 32-Core CPU with 263 GB of RAM. 
The operating system is Ubuntu 20.04 LTS with PyTorch 2.0.1. We utilize this server to simulate an MASP and multiple uniformed distributed mobile users.
\begin{figure*}[tbp]
\centerline{\includegraphics[width=1.9\columnwidth]{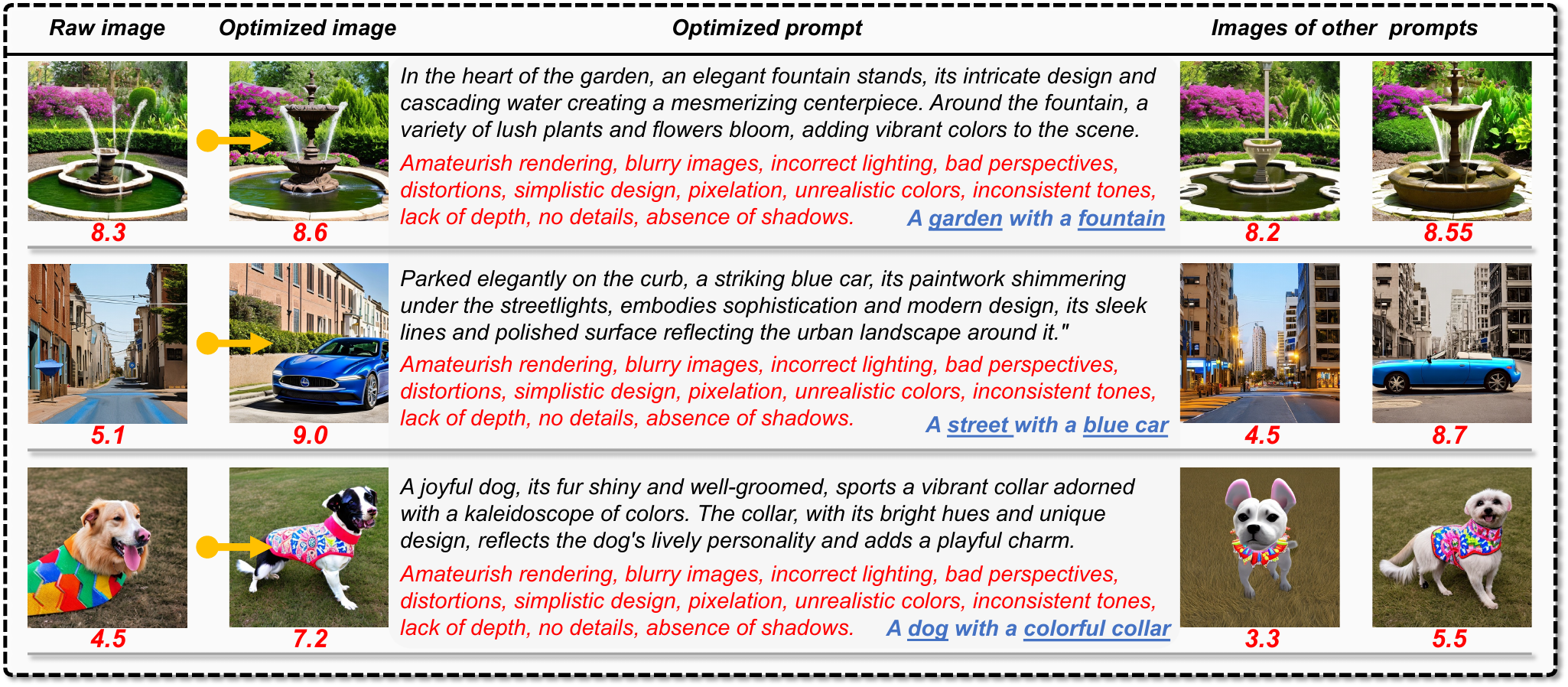}}
\caption{The effectiveness of interactive prompt engineering. Note that these cases show that prompt engineering cannot always improve generation quality. For instance, in the second row, the image generated by the refined prompt also fails to illustrate the blue car.}
\label{prompt}
\vspace{-0.5cm}
\end{figure*}
\begin{figure}[tbp]
\centerline{\includegraphics[width=0.95\columnwidth]{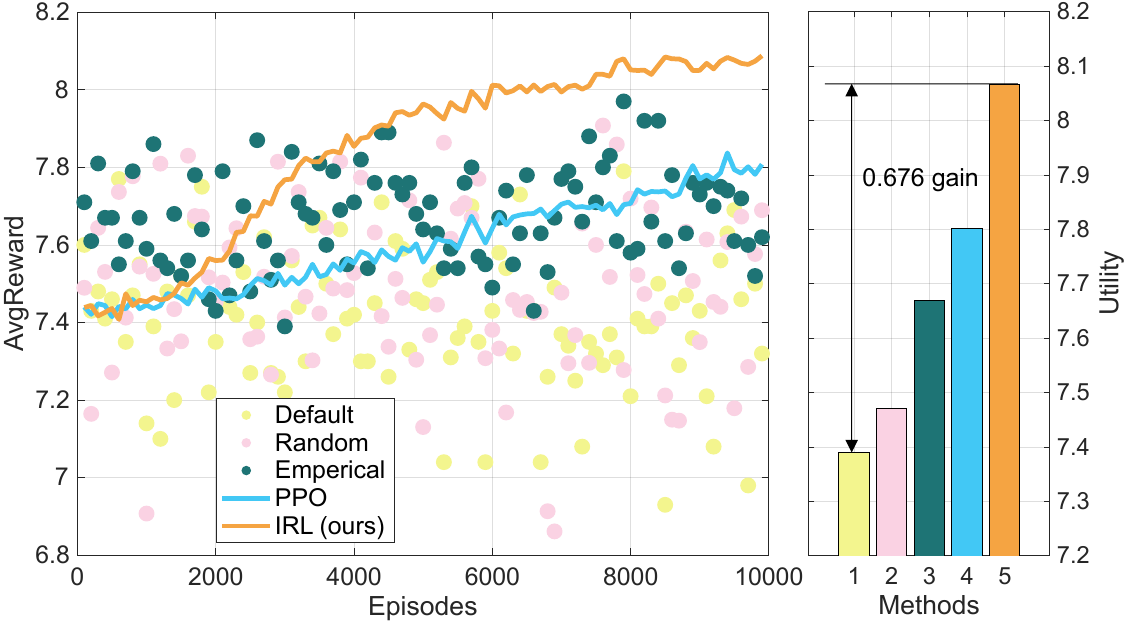}}
\caption{The training curves and converged utilities of default (i.e., without prompt engineering), random, empirical, PPO, and IRL prompt engineering policies.}
\label{gail}
\vspace{-0.5cm}
\end{figure}

\textbf{Configurations.} We equip an MASP with Stable Diffusion v2.0 \cite{SDpaper} to realize the text-to-image AIGC services. The diffusion step is set to 25. The user prompts are generated by ChatGPT (empowered by the GPT-4 model) in the form of ``\texttt{A [A], with [B]}". The demonstration prompts are randomly sampled from the user prompts. Based on \cite{YQNetwork}, we consider six aspects for refining raw prompts, namely \textit{object description}, \textit{environment}, \textit{mood}, \textit{lighting}, \textit{quality booster}, and \textit{negative effects}. 
\begin{itemize}
    \item \textbf{Object Description}: To facilitate fine-grained image generation, detailed descriptions of \texttt{[a]} and \texttt{[b]}'s type, texture, and features should be provided. Such details enable the AIGC model to associate more pre-learned knowledge, resulting in delicate images.
    \item \textbf{Environment}: The environment fills the background of the image, creating a real, harmonious, and beautiful scene for \texttt{[b]}. Furthermore, environment description can prevent the AIGC model from only searching and stacking the found materials about \texttt{[a]} and \texttt{[b]}, thereby further enhancing the composition quality.
    \item \textbf{Mood}: Mood describes the emotion that the users intend to convey through the image, such as joy, sadness, or hesitation, which is reflected by the color palette, the facial expressions of the characters, etc.
    \item \textbf{Lighting}: Lighting is a fundamental factor in determining the texture and authenticity of AI-generated images. The prompt for lighting should clarify the light sources and the effect of light shining on different objects.
    \item \textbf{Quality Booster}: Quality boosters refer to various adjectives that describe the user desirability, e.g., \textit{high-quality}, \textit{2k resolution}, and \textit{real texture}. By sampling from the distribution of high-quality images, the newly generated images tend to acquire higher aesthetic quality.
    \item \textbf{Negative Effects}: Negative effects depict situations that might decrease image quality. By moving the sampling distribution away from data distributions containing such negative effects, the AIGC model can prevent generated images from containing effects that decrease the quality or are undesired by users.
\end{itemize}
Then, seven prompt engineering strategies are presented (see TABLE II). This aims to filter out some irrational arrangements and reduce the action space, thereby improving the training efficiency of D$^3$PG. Note that such a principle is widely adopted since human experience and knowledge play an important role in prompt engineering \cite{2309.08532}. Moreover, users are free to customize prompt enriching aspects and prompt engineering strategies when applying our proposal to their applications. The detailed experimental settings are summarized in TABLE III.


\subsection{Rationale of Assessing Agent}
First, we investigate the rationale of the LLM-empowered assessing agent, i.e., whether it can assess the given image fairly and comprehensively.
Fig. \ref{rationale} shows the assessment of a series of images using three methods, namely our assessing agent, NIMA \cite{8352823}, and Image-reward \cite{10.5555/3666122.3666822}.
Note that NIMA is a classic and widely adopted aesthetic quality metric trained on large-scale human feedback.
Image-reward is one of the latest AIGC-oriented assessing frameworks, which utilizes BLIP as the backbone model and supports multimodal understanding (i.e., can check the alignment between image content and prompt).
Similarly to NIMA, image-reward is also trained on large-scale human annotations, where images are rated from three aspects, namely alignment, fidelity, and harmlessness.
From Fig. \ref{rationale}, we can observe that our method outperforms NIMA and image-reward in three dimensions.
First, without multimodal understanding ability, various existing assessment methods, such as NIMA, BRISQUE\footnote{https://pypi.org/project/brisque/}, and LPIPS\footnote{https://pypi.org/project/lpips/}, cannot fit the AIGC scenarios.
The reason is that AIGC generations usually involve modality transfers, e.g., generating images from texts.
As marked by \circled{1} in Fig. \ref{rationale}, NIMA cannot associate the image with its textual prompt and gives a high score to an image that fails to illustrate the \texttt{blue car}.
Second, attributed to the massive knowledge of LLM, the assessing agent can better simulate real humans and understand the image semantics more precisely.
For instance, it correctly identifies fog in the forest, while other methods misjudge it as blurs and give low scores (see \circled{1} in Fig. \ref{rationale}).
Finally, our assessing agent can explain the reasons behind the scoring, which greatly outperforms conventional methods whose results are unexplainable.
In the above example, precise and rational explanations of the fog are provided (see \circled{3} in Fig. \ref{rationale}).

After the above analysis, we investigate whether the assessing agent's scores are consistent with aesthetics. 
To do so, we randomly select 140 images and arrange them in order from low to high scores. 
Afterward, we extract their image-reward scores as references and perform a curve fitting. 
From Fig. \ref{rationale}, we can conclude that the assessing agent and image-reward maintain high-level alignment in terms of aesthetic judgment.
Since the latter is a widely adopted and well-proven aesthetics assessment metric for AIGC, the rationale of our assessing agent is validated.

\begin{figure*}[tbp]
\centerline{\includegraphics[width=2\columnwidth]{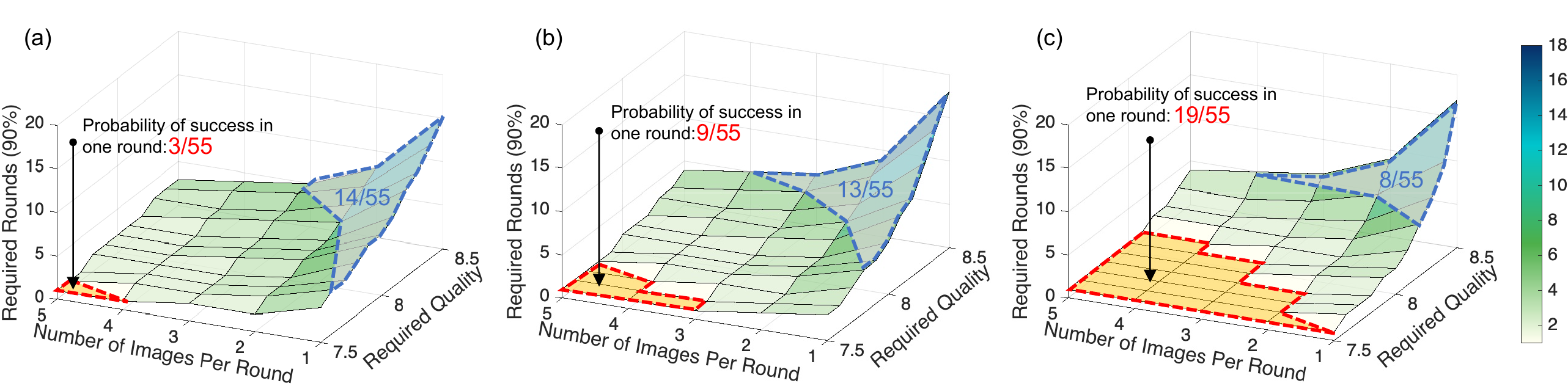}}
\caption{The number of required service rounds with respect to varying user requirements and inference numbers per round. (a): Default; (b): Empirical; (c): Our IRL-based approach. The orange and blue zones highlight the conditions in which only one and more than five rounds are required, respectively.}
\label{number}
\vspace{-0.3cm}
\end{figure*}

\begin{figure}[tbp]
\centerline{\includegraphics[width=0.95\columnwidth]{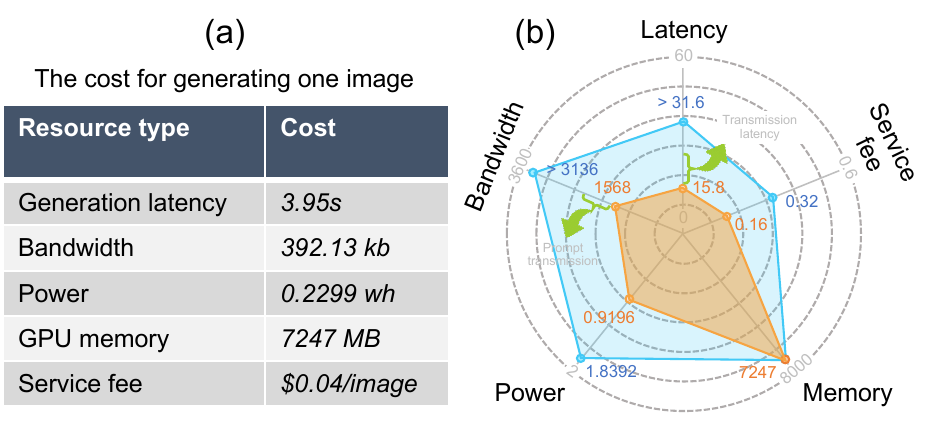}}
\caption{(a): The resource consumption for generating each image. (b): The resource consumption for performing one and two rounds of service (suppose four images are generated in each round).}
\label{hardware}
\vspace{-0.3cm}
\end{figure}

\subsection{Inspection on Prompt Engineering Policy}
In this part, we evaluate the efficiency of $\pi^{(p)}_\omega$ through two comprehensive studies.
First, Fig. \ref{prompt} illustrates the effectiveness of prompt engineering in improving generation quality. 
We randomly select three raw prompts, perform all types of prompt engineering strategies shown in TABLE II, and evaluate the generation quality using our assessing agent. 
The results clearly demonstrate that the images generated by the raw prompts suffer from significant flaws.
For instance, the water flow and fountain base are misaligned, and the blue car and dog's legs are missing from the scene. 
By enriching the prompts, the AIGC model achieves richer task descriptions and instructions, leading to substantial improvements in prompt alignment, object rendering, and image composition. 
Quantitative scores validate these improvements. 
Particularly, we observe that \textit{Strategy 6} consistently leads to the optimal generation quality across all test cases.

Then, we train $\pi^{(p)}_\omega$ using the demonstration dataset. 
To prove the superiority of our proposal in policy imitation with small-scale datasets, we set PPO as the baseline. 
Note that PPO maintains the same network architecture for policy refinement and action evaluation, while our IRL-based approach introduces a discriminator and follows an adversarial training paradigm. 
Additionally, we implement two non-learning baselines: random and empirical (i.e., always selecting \textit{Strategy 6}). 
As shown in Fig. \ref{gail}, the random policy performs similarly to non-prompt engineering. 
The empirical policy achieves higher rewards and smaller variance, as empirical experience ensures that the optimal/near-optimal policy can be selected in many cases. 
Through policy reinforcement, PPO demonstrates better adaptability and achieves more stable improvements compared to non-learning baselines but faces limitations in two aspects: 1) PPO relies solely on reward signals for optimizing policy, which can be insufficient when learning complex prompt engineering strategies from limited demonstrations; 2) The direct policy optimization in PPO may not effectively capture the nuanced relationships between prompts and generation quality present in expert demonstrations.
In contrast, our IRL approach adopts an adversarial training paradigm, which provides several key advantages: 1) The discriminator learns to distinguish between expert and policy behaviors, providing a more informative learning signal than pure reward values; 2) The adversarial training allows for better imitation of expert prompt engineering strategies by capturing both the actions and their underlying patterns; 3) The generator-discriminator architecture is particularly effective with limited demonstration data, as it can generalize from few examples through the adversarial learning process. Consequently, our IRL approach achieves the best efficiency in selecting optimal prompt engineering strategies according to specific user requests, showing consistent improvement throughout training and reaching the highest utility of approximately 8.06.

\subsection{Impact of Generation Quality on Mobile-edge Networks}
The increased generation quality directly leads to fewer re-generations, which saves substantial networking resources. 
To quantify this benefit, we explore the required number of service rounds under varying user quality requirements and MASP's per-round inference numbers. 
Specifically, we evaluate scenarios where user-required quality ranges from 7.5 to 8.5, and the number of images generated per round varies from 1 to 5. 
We compare three representative prompt engineering strategies, namely default, empirical (i.e., always selecting \textit{Strategy 6}), and IRL.
Suppose that the generation quality of each strategy follows a standard distribution. 
The mean and variance can be fitted from the sample results. 
Setting the confidence level at 90\%, we calculate the required number of service rounds. 
As shown in Fig. \ref{number}, without prompt engineering, the probability of zero re-generation is only 3/55. The empirical prompt engineering strategy improves the probability of single-round success to 9/55.
In contrast, our IRL-based approach significantly outperforms both baselines, outperforming none and empirical strategies by 6.3$\times$ and 2.1$\times$, respectively. Moreover, the probability of requiring more than five service rounds (indicated by the blue regions) is significantly reduced.

Fig. \ref{hardware}(a) benchmarks the consumption of five critical resources required to generate one image, namely generation latency, bandwidth, power, GPU memory, and service fee\footnote{The reference fee can be found at https://openai.com/api/pricing/}. 
These measurements reveal substantial resource demands of AIGC inferences for mobile servers. 
Furthermore, when one re-generation is performed, the resource overhead more than doubles since the refinement and re-transmission of prompts consume additional time and bandwidth (as shown in Fig. \ref{hardware}(b)).
Beyond quantifiable resource costs, failed generation attempts also negatively impact user QoE as their service requests remain unfilled. 
Contributed to improved generation quality through prompt IRL-based engineering, the proposed intelligent mobile AIGC service scheme achieves significantly higher resource efficiency.

\begin{figure}[tbp]
\centerline{\includegraphics[width=0.95\columnwidth]{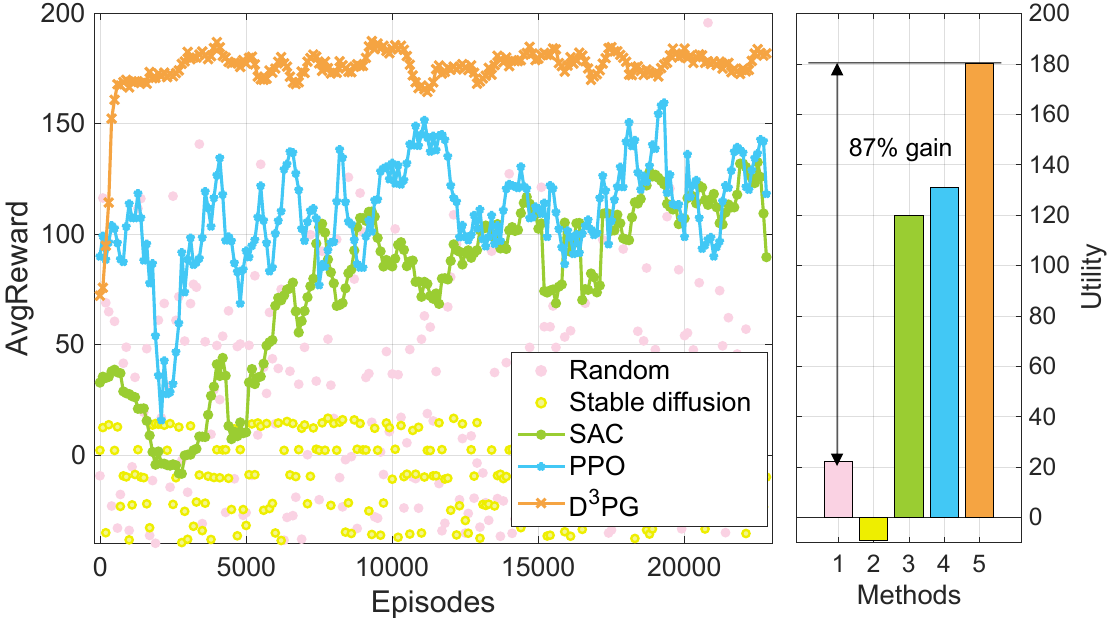}}
\caption{The training curves and converged utilities of random, Stable Diffusion, SAC, PPO, and D$^3$PG for mobile AIGC service provisioning.}
\label{d3pg}
\vspace{-0.3cm}
\end{figure}

\begin{figure}[tbp]
\centerline{\includegraphics[width=0.95\columnwidth]{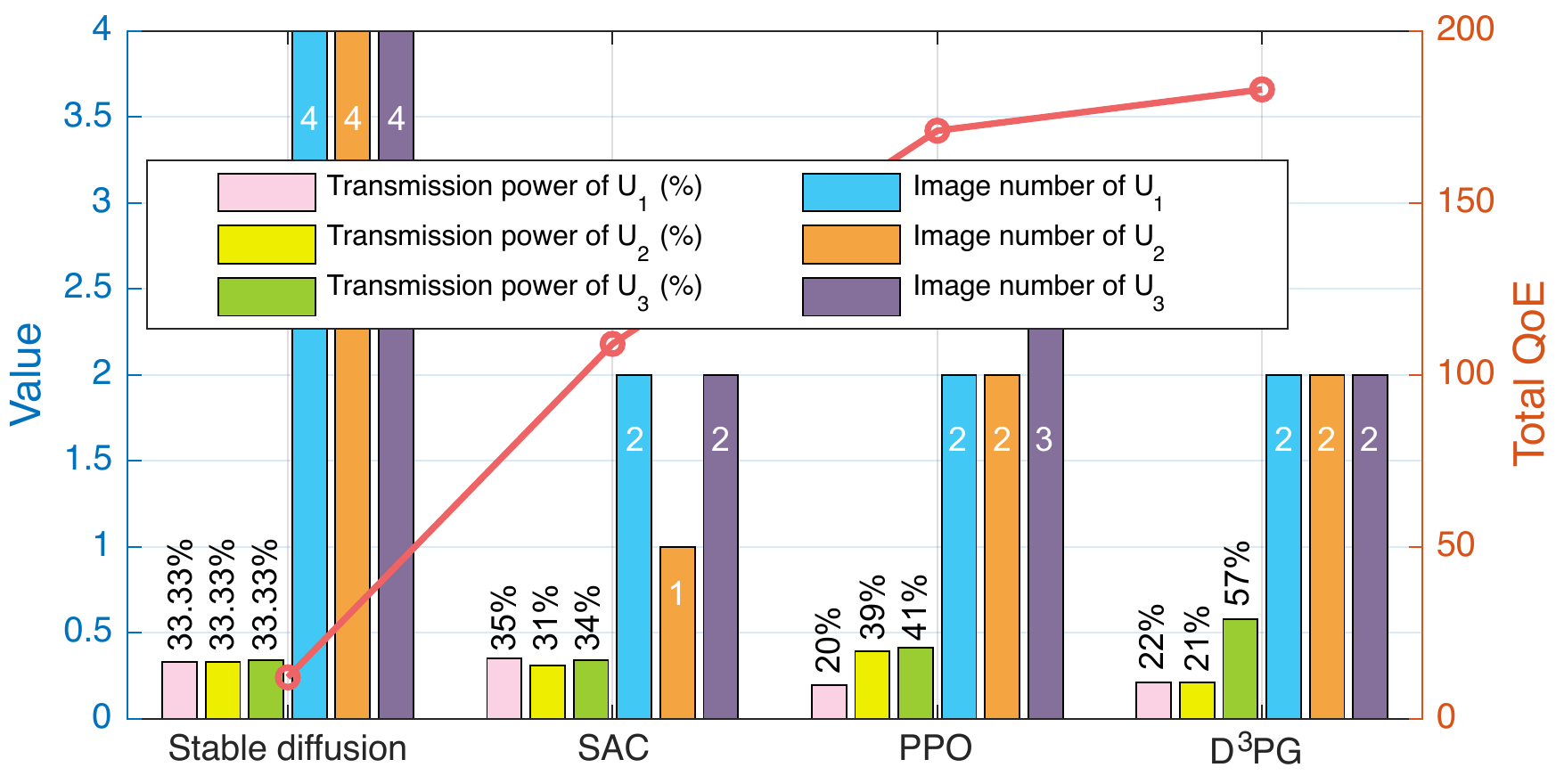}}
\caption{The exemplar service provisioning shame generated by four different methods and resulting QoE values.}
\label{d3pg-utility}
\vspace{-0.3cm}
\end{figure}

\subsection{Evaluation of Service Provisioning Policy}
Our interactive prompt engineering maximizes the generation quality of each inference trial. 
In this part, we optimize the number of inference trials per round and transmission power allocation to further improve user QoE. 
Fig. \ref{d3pg} illustrates the training curves and converged utility of different methods. 
Apart from practical solutions like Stable Diffusion, we employ two representative DRL-based baselines, namely PPO and Soft Actor-Critic (SAC) \cite{10638833}.
We observe that Stable Diffusion performs poorly as static service provisioning cannot meet heterogeneous user requirements. 
Specifically, users with simpler tasks receive excessive resources, while those with complex tasks receive insufficient support, leading to resource inefficiency. 
The random approach occasionally achieves satisfactory rewards but suffers from high variance, as shown by the scattered pink dots.
Learning-based methods achieve better performance by adapting service provisioning to different user requirements. 
As illustrated in Fig. \ref{d3pg}, both PPO and SAC improve over episodes, with PPO demonstrating faster initial learning while SAC achieving more stable long-term performance. 
Finally, D$^3$PG significantly outperforms both baselines, achieving at most 87\% improvement in converged utility.
This superiority can be attributed to two factors. First, integrating diffusion models into the actor-network enhances environmental exploration by providing structured noise injection, allowing D$^3$PG to discover better policies in the complex action space.
Second, compared to the fixed Gaussian noise in PPO and SAC, our diffusion-based policy refinement enables more precise adjustment of the action distribution, leading to better convergence and more robust performance.

Finally, Fig. \ref{d3pg-utility} shows the decisions of four methods when the user prompts are ``\texttt{A dog with a colorful collar}", ``\texttt{A garden with a fountain}", ``\texttt{A city with blue car}" and the quality thresholds are 7.6, 8.2, and 8.5, respectively.
We can observe that Stable diffusion adopts a fixed strategy with uniform transmission power allocation (i.e., 33.33\% for each user) and four inference trials each round, resulting in inefficient resource utilization. 
All three learning-based methods generate customized service provisioning schemes.
Due to inefficient environment exploration and policy refinement, SAC allocates transmission power nearly equally, leading to insufficient resources for users with higher quality thresholds. 
In contrast, PPO and D$^3$PG demonstrate superior capability in dynamic resource allocation. 
PPO adjusts both the transmission power distribution (i.e., 20-41\%) and inference trials, while D$^3$PG achieves the most efficient allocation by assigning significantly higher transmission power (i.e., 57\% of $P_{\mathrm{total}}$) to the most demanding user while maintaining balanced inference trials. 
Accordingly, D$^3$PG achieves the highest overall QoE, with 67.8\% and 7.0\% improvements over SAC and PPO, respectively.

\vspace{-0.2cm}
\section{Conclusion}
In this paper, we have presented an intelligent mobile AIGC service scheme with interactive prompt engineering and dynamic service provisioning. 
Specifically, to increase AIGC generation quality, we have proposed an IRL-based approach that leverages demonstration datasets and policy imitation to acquire optimal prompt engineering strategies. 
Then, different from fixed service provisioning, we have formulated the QoE optimization problem with respect to wireless transmission power and the number of AIGC inference trials.
Furthermore, we have presented the D$^3$PG algorithm for QoE optimization, which integrates diffusion models into the DRL framework to enhance environmental exploration capabilities.
Extensive numerical results have validated that our proposals effectively improve generation quality and user QoE through reduced service rounds and optimized resource allocation.
More importantly, our proposals are unified and can support various mobile AIGC applications.



\ifCLASSOPTIONcaptionsoff
  \newpage
\fi

\bibliographystyle{IEEEtran}
\bibliography{InteractivePE}

\begin{thebibliography}{10}
\providecommand{\url}[1]{#1}
\csname url@samestyle\endcsname
\providecommand{\newblock}{\relax}
\providecommand{\bibinfo}[2]{#2}
\providecommand{\BIBentrySTDinterwordspacing}{\spaceskip=0pt\relax}
\providecommand{\BIBentryALTinterwordstretchfactor}{4}
\providecommand{\BIBentryALTinterwordspacing}{\spaceskip=\fontdimen2\font plus
\BIBentryALTinterwordstretchfactor\fontdimen3\font minus \fontdimen4\font\relax}
\providecommand{\BIBforeignlanguage}[2]{{%
\expandafter\ifx\csname l@#1\endcsname\relax
\typeout{** WARNING: IEEEtran.bst: No hyphenation pattern has been}%
\typeout{** loaded for the language `#1'. Using the pattern for}%
\typeout{** the default language instead.}%
\else
\language=\csname l@#1\endcsname
\fi
#2}}
\providecommand{\BIBdecl}{\relax}
\BIBdecl

\bibitem{10398474}
Y.~Zhang, J.~Zhang, S.~Yue, W.~Lu, J.~Ren, and X.~Shen, ``Mobile generative ai: Opportunities and challenges,'' \emph{IEEE Wirel. Commun.}, vol.~31, no.~4, pp. 58--64, 2024.

\bibitem{DuAIGC}
H.~Du \emph{et~al.}, ``Enabling {AI}-generated content {(AIGC)} services in wireless edge networks,'' \emph{IEEE Wirel. Commun.}, vol.~31, no.~3, pp. 226--234, 2024.

\bibitem{YQNetwork}
Y.~Liu \emph{et~al.}, ``Optimizing mobile-edge {AI}-generated everything {(AIGX)} services by prompt engineering: Fundamental, framework, and case study,'' \emph{IEEE Netw.}, vol.~38, no.~5, pp. 220--228, 2024.

\bibitem{GPT-3Cost}
\BIBentryALTinterwordspacing
The training costs of {GPT}-3. 2024. [Online]. Available: \url{https://lambdalabs.com/blog/demystifying-gpt-3}
\BIBentrySTDinterwordspacing

\bibitem{Snapdragon}
\BIBentryALTinterwordspacing
The introduction to {Q}ualcomm snapdragon chip. 2024. [Online]. Available: \url{https://www.qualcomm.com/snapdragon/overview}
\BIBentrySTDinterwordspacing

\bibitem{10628024}
J.~Zhang, Z.~Wei, B.~Liu, X.~Wang, Y.~Yu, and R.~Zhang, ``Cloud-edge-terminal collaborative aigc for autonomous driving,'' \emph{IEEE Wirel. Commun.}, vol.~31, no.~4, pp. 40--47, 2024.

\bibitem{qualcomm}
\BIBentryALTinterwordspacing
The word's first on-device {S}table {D}iffusion version by {Q}ualcomm. 2023. [Online]. Available: \url{https://www.qualcomm.com/news/onq/2023/02/worlds-first-on-device-demonstration-of-stable-diffusion-on-android}
\BIBentrySTDinterwordspacing

\bibitem{knowledge}
T.~Salimans and J.~Ho, ``Progressive distillation for fast sampling of diffusion models,'' in \emph{Proc. ICLR}, 2022, pp. 1--21.

\bibitem{chen2023speed}
Y.-H. Chen \emph{et~al.}, ``Speed is all you need: On-device acceleration of large diffusion models via {GPU}-aware optimizations,'' in \emph{Proc. CVPR Workshops}, 2023, pp. 4650--4654.

\bibitem{xu2023sparks}
M.~Xu \emph{et~al.}, ``Sparks of {GPTs} in edge intelligence for metaverse: Caching and inference for mobile {AIGC} services,'' 2023.

\bibitem{10172151}
H.~Du \emph{et~al.}, ``Exploring collaborative distributed diffusion-based {AI}-generated content {(AIGC)} in wireless networks,'' \emph{IEEE Netw.}, vol.~38, no.~3, pp. 178--186, 2024.

\bibitem{10233667}
J.~Wen \emph{et~al.}, ``Freshness-aware incentive mechanism for mobile {AI}-generated content ({AIGC}) networks,'' in \emph{Proc. ICCC}, 2023, pp. 1--6.

\bibitem{Complexity}
S.~Mishra, D.~Z. Chen, and X.~S. Hu, ``Image complexity guided network compression for biomedical image segmentation,'' \emph{ACM J. Emerg. Technol. Comput. Syst.}, vol.~18, no.~2, pp. 1--23, 2021.

\bibitem{10.1145/3560815}
P.~Liu, W.~Yuan, J.~Fu, Z.~Jiang, H.~Hayashi, and G.~Neubig, ``Pre-train, prompt, and predict: A systematic survey of prompting methods in natural language processing,'' \emph{ACM Comput. Surv.}, vol.~55, no.~9, pp. 1--35, 2023.

\bibitem{10.1145/3641289}
Y.~Chang \emph{et~al.}, ``A survey on evaluation of large language models,'' \emph{ACM Trans. Intell. Syst. Technol.}, vol.~15, no.~3, pp. 1--45, 2024.

\bibitem{Prompt-OIRL}
H.~Sun, A.~Huyuk, and M.~Schaar, ``Query-dependent prompt evaluation and optimization with offline inverse {RL},'' in \emph{Proc. ICLR}, 2024, pp. 1--56.

\bibitem{zhang2023tempera}
T.~Zhang, X.~Wang, D.~Zhou, D.~Schuurmans, and J.~E. Gonzalez, ``{TEMPERA}: Test-time prompt editing via reinforcement learning,'' in \emph{Proc. ICLR}, 2023, pp. 1--16.

\bibitem{10096710}
J.~Chen, X.~Song, Z.~Peng, B.~Zhang, F.~Pan, and Z.~Wu, ``Lightgrad: Lightweight diffusion probabilistic model for text-to-speech,'' in \emph{Proc. ICASSP}, 2023, pp. 1--5.

\bibitem{li2023diffnas}
W.~Li, X.~Su, S.~You, F.~Wang, C.~Qian, and C.~Xu, ``Diffnas: Bootstrapping diffusion models by prompting for better architectures,'' \emph{ArXiv preprint: ArXiv:2310.04750}, 2023.

\bibitem{li2023snapfusion}
Y.~Li \emph{et~al.}, ``Snapfusion: Text-to-image diffusion model on mobile devices within two seconds,'' in \emph{Proc. NeurIPS}, 2023, pp. 1--17.

\bibitem{10398264}
X.~Huang \emph{et~al.}, ``Federated learning-empowered {AI}-generated content in wireless networks,'' \emph{IEEE Netw.}, vol.~38, no.~5, pp. 304--313, 2024.

\bibitem{AIGCSemCom}
R.~Cheng, Y.~Sun, D.~Niyato, L.~Zhang, L.~Zhang, and M.~Imran, ``A wireless {AI}-generated content ({AIGC}) provisioning framework empowered by semantic communication,'' \emph{IEEE Trans. Mob. Comput.}, pp. 1--14, 2024.

\bibitem{MATTING}
Y.~Zhang \emph{et~al.}, ``Matting moments: A unified data-driven matting engine for mobile aigc in photo gallery,'' in \emph{Proc. IJCAI}, 2023, pp. 7183--7186.

\bibitem{wen2023hard}
Y.~Wen, N.~Jain, J.~Kirchenbauer, M.~Goldblum, J.~Geiping, and T.~Goldstein, ``Hard prompts made easy: Gradient-based discrete optimization for prompt tuning and discovery,'' in \emph{Proc. NeurIPS}, 2023, pp. 1--18.

\bibitem{10210127}
T.~Guo, S.~Guo, J.~Wang, X.~Tang, and W.~Xu, ``Promptfl: Let federated participants cooperatively learn prompts instead of models - federated learning in age of foundation model,'' \emph{IEEE Trans. Mob. Comput.}, vol.~23, no.~5, pp. 5179--5194, 2024.

\bibitem{ACL}
R.~Pryzant, D.~Iter, J.~Li, Y.~T. Lee, C.~Zhu, and M.~Zeng, ``Automatic prompt optimization with ``gradient descent" and beam search,'' in \emph{Proc. EMNLP}, 2023, pp. 7957--7968.

\bibitem{2309.08532}
Q.~Guo \emph{et~al.}, ``Connecting large language models with evolutionary algorithms yields powerful prompt optimizers,'' in \emph{Proc. ICLR}, 2014, pp. 1--24.

\bibitem{Evoprompting}
A.~Chen, D.~Dohan, and D.~So, ``Evoprompting: Language models for code-level neural architecture search,'' in \emph{Proc. NeurIPS}, 2023, pp. 7787 -- 7817.

\bibitem{deng-etal-2022-rlprompt}
M.~Deng \emph{et~al.}, ``{RLP}rompt: Optimizing discrete text prompts with reinforcement learning,'' in \emph{Proc. EMNLP}, 2022, pp. 3369--3391.

\bibitem{9186847}
T.~Li, Y.~Li, M.~A. Hoque, T.~Xia, S.~Tarkoma, and P.~Hui, ``To what extent we repeat ourselves? {D}iscovering daily activity patterns across mobile app usage,'' \emph{IEEE Trans. Mob. Comput.}, vol.~21, no.~4, pp. 1492--1507, 2022.

\bibitem{du2023usercentric}
H.~Du \emph{et~al.}, ``User-centric interactive {AI} for distributed diffusion model-based {AI}-generated content,'' \emph{ArXiv preprint: ArXiv:2311.11094}, 2023.

\bibitem{9044870}
H.~Du, J.~Zhang, J.~Cheng, and B.~Ai, ``Sum of fisher-snedecor f random variables and its applications,'' \emph{IEEE Open J. Commun. Soc.}, vol.~1, pp. 342--356, 2020.

\bibitem{5654629}
M.~A. Rahman and H.~Harada, ``New exact closed-form pdf of the sum of nakagami-m random variables with applications,'' \emph{IEEE Trans. Commun.}, vol.~59, no.~2, pp. 395--401, 2011.

\bibitem{655405}
L.~Miller and J.~Lee, ``Ber expressions for differentially detected /spl pi//4 dqpsk modulation,'' \emph{IEEE Trans. Commun.}, vol.~46, no.~1, pp. 71--81, 1998.

\bibitem{OpenCLIP}
M.~Cherti \emph{et~al.}, ``Reproducible scaling laws for contrastive language-image learning,'' in \emph{Proc. CVPR}, 2023, pp. 2818--2829.

\bibitem{PicScore}
P.~C. Neto, A.~F. Sequeira, J.~S. Cardoso, and P.~Terhörst, ``Pic-score: Probabilistic interpretable comparison score for optimal matching confidence in single- and multi-biometric (face) recognition,'' in \emph{Proc. CVPR}, 2023, pp. 1021--1029.

\bibitem{roleprompting}
A.~Kong \emph{et~al.}, ``Better zero-shot reasoning with role-play prompting,'' in \emph{Proc. NAACL}, 2024, pp. 1--15.

\bibitem{CLIP}
A.~Radford \emph{et~al.}, ``Learning transferable visual models from natural language supervision,'' in \emph{Proceedings of the 38th International Conference on Machine Learning}, 2021, pp. 8748--8763.

\bibitem{LangChain}
\BIBentryALTinterwordspacing
The introduction to {L}ang{C}hain. 2024. [Online]. Available: \url{https://www.langchain.com/}
\BIBentrySTDinterwordspacing

\bibitem{MemGPT}
\BIBentryALTinterwordspacing
The introduction to {M}em{GPT}. 2024. [Online]. Available: \url{https://memgpt.ai/}
\BIBentrySTDinterwordspacing

\bibitem{GAIL}
J.~Ho and S.~Ermon, ``Generative adversarial imitation learning,'' in \emph{Proc. NeurIPS}, 2016, pp. 4572 -- 4580.

\bibitem{10032267}
R.~Zhang, K.~Xiong, Y.~Lu, P.~Fan, D.~W.~K. Ng, and K.~B. Letaief, ``Energy efficiency maximization in {RIS}-assisted {SWIPT} networks with {RSMA}: A {PPO}-based approach,'' \emph{IEEE J. Sel. Areas Commun.}, vol.~41, no.~5, pp. 1413--1430, 2023.

\bibitem{6263849}
T.~Hossfeld, S.~Egger, R.~Schatz, M.~Fiedler, K.~Masuch, and C.~Lorentzen, ``Initial delay vs. interruptions: Between the devil and the deep blue sea,'' in \emph{Proc. QoMEX}, 2012, pp. 1--6.

\bibitem{WF_Law}
\BIBentryALTinterwordspacing
The introduction to weber-fechner law. 2024. [Online]. Available: \url{https://www.raggeduniversity.co.uk/wp-content/uploads/2018/03/Weber-Fechner-Law.pdf}
\BIBentrySTDinterwordspacing

\bibitem{DiffusionDRL1}
H.~Du \emph{et~al.}, ``Diffusion-based reinforcement learning for edge-enabled ai-generated content services,'' \emph{IEEE Trans. Mob. Comput.}, vol.~23, no.~9, pp. 8902--8918, 2024.

\bibitem{zhu2023diffusion}
Z.~Zhu \emph{et~al.}, ``Diffusion models for reinforcement learning: A survey,'' \emph{arXiv preprint arXiv:2311.01223}, 2023.

\bibitem{DDPM}
J.~Ho, A.~Jain, and P.~Abbeel, ``Denoising diffusion probabilistic models,'' in \emph{Proc. NeurIPS}, 2023, pp. 6840 -- 6851.

\bibitem{SDpaper}
\BIBentryALTinterwordspacing
Stable diffusion model. 2023. [Online]. Available: \url{https://stability.ai/blog/stable-diffusion-public-release}
\BIBentrySTDinterwordspacing

\bibitem{8352823}
H.~Talebi and P.~Milanfar, ``{NIMA}: Neural image assessment,'' \emph{IEEE Trans. Image Process.}, vol.~27, no.~8, pp. 3998--4011, 2018.

\bibitem{10.5555/3666122.3666822}
J.~Xu \emph{et~al.}, ``Imagereward: learning and evaluating human preferences for text-to-image generation,'' in \emph{Proc. NeurIPS}, 2023, pp. 15\,903--15\,935.

\bibitem{10638833}
Y.~Liang, H.~Tang, H.~Wu, Y.~Wang, and P.~Jiao, ``Lyapunov-guided offloading optimization based on soft actor-critic for isac-aided internet of vehicles,'' \emph{IEEE Trans. Mob. Comput.}, vol.~23, no.~12, pp. 14\,708--14\,721, 2024.

\end{thebibliography}
\vfill

\end{document}